\RequirePackage{fix-cm}
\documentclass[twocolumn,epjc3]{svjour3}  
\smartqed  
\RequirePackage{graphicx}
\journalname{Eur. Phys. J. C}
\usepackage{epsfig}
\usepackage{amsfonts}
\usepackage{amsmath}
\usepackage{amssymb}
\usepackage{dsfont}
\usepackage{xcolor}         
\usepackage[colorlinks,citecolor=red,urlcolor=blue,bookmarks=false,hypertexnames=true]{hyperref}             
\usepackage{hyperref}
\usepackage{color}
\usepackage{cite}
\usepackage{braket}	
\usepackage{caption}
\usepackage{subcaption}	
\usepackage{csquotes}
\usepackage{tikz}
\usepackage{widetext}

\usepackage{feynmp-auto}

\RequirePackage[colorinlistoftodos,prependcaption,textsize=tiny]{todonotes}
\usepackage{soul}

\newcommand{\al}{\alpha}

\newcommand{\Fp}{F_\pi}

\newcommand{\mpi}{m_{\pi}}

\newcommand{\dslash}[1]{#1 \llap{/\kern-0.5pt}}
\newcommand{\Dslash}[1]{#1 \llap{/\kern+1.5pt}}
\newcommand{\DDslash}[1]{#1 \llap{/\kern+2.3pt}}
\newcommand{\dslashh}[1]{#1 \llap{/\kern+1pt}}
\newcommand{\abs}[1]{\left|#1\right|}

\newcommand{\boldq}{\mbox{\boldmath $q$}}
\newcommand{\boldQ}{\mbox{\boldmath $Q$}}
\newcommand{\boldp}{\mbox{\boldmath $p$}}
\newcommand{\boldk}{\mbox{\boldmath $k$}}

\newcommand{\boldsigma}{\mbox{\boldmath $\sigma$}}

\newcommand{\bea}{\begin{eqnarray}}
\newcommand{\eea}{\end{eqnarray}}
\newcommand{\bee}{\begin{equation}}
\newcommand{\ee}{\end{equation}}
\newcommand{\bma}{\begin{pmatrix}}
\newcommand{\ema}{\end{pmatrix}}
\newcommand{\nn}{\nonumber}

\renewcommand{\vec}[1]{\boldsymbol{#1}}

\newcommand{\CG}[6]{\left\langle #1 #4, #2 #5 \middle\vert #3 #6 \right\rangle}
\newcommand{\mc}{m_{\chi}}

\newcommand{\diff}[1]{\text{d}{#1} \, }
\newcommand{\MeV}{\mbox{ MeV}}

\newcommand{\fm}{\mbox{ fm}}

\newcommand{\chipt}{$\chi$PT}

\newcommand{\Fis}[1]{\mathcal{F}_{\mathrm{is}}^{(#1)}(\boldq^2)}
\newcommand{\nlo}[1]{N${}^{#1}$LO}
\newcommand{\Fisa}[2]{\mathcal{F}_{\mathrm{is},#2}^{(#1)}(\boldq^2)}

\newcommand{\dd}{\mathrm{d}}

\newcommand{\lab}{l_{12}}
\newcommand{\lpab}{l_{12}^{\prime}}
\newcommand{\mab}{m_{12}}
\newcommand{\mpab}{m_{12}^{\prime}}
\newcommand{\jab}{j_{12}}
\newcommand{\jpab}{j_{12}^{\prime}}
\newcommand{\sab}{s_{12}}
\newcommand{\spab}{s_{12}^{\prime}}
\newcommand{\mtab}{m^{t}_{12}}
\newcommand{\mtpab}{m_{12}^{t \prime}}
\newcommand{\msab}{m^{s}_{12}}
\newcommand{\mspab}{m_{12}^{s \prime}}
\newcommand{\tab}{t_{12}}
\newcommand{\tpab}{t_{12}^{\prime}}

\newcommand{\lthreec}{l_{3}}
\newcommand{\lthreepc}{l_{3}^{\prime}}

\newcommand{\Ithreec}{I_{3}}
\newcommand{\Ithreepc}{I_{3}^{\prime}}
\newcommand{\sthreec}{s_{3}}
\newcommand{\sthreepc}{s_{3}^{\prime}}

\newcommand{\tthreec}{t_{3}}
\newcommand{\tthreepc}{t_{3}^{\prime}}

\newcommand{\lfourc}{l_{4}}
\newcommand{\lfourpc}{l_{4}^{\prime}}
\newcommand{\mfourc}{m_{4}}
\newcommand{\mfourpc}{m_{4}^{\prime}}
\newcommand{\Ifourc}{I_{4}}
\newcommand{\Ifourpc}{I_{4}^{\prime}}
\newcommand{\sfourc}{s_{4}}
\newcommand{\sfourpc}{s_{4}^{\prime}}
\newcommand{\mtfourc}{m^{t}_{4}}
\newcommand{\mtfourpc}{m_{4}^{t \prime}}
\newcommand{\msfourc}{m^{s}_{4}}
\newcommand{\msfourpc}{m_{4}^{s \prime}}
\newcommand{\tfourc}{t_{4}}
\newcommand{\tfourpc}{t_{4}^{\prime}}
\newcommand{\mlfourc}{m_{4}^{l}}
\newcommand{\mlfourpc}{m_{4}^{l \prime}}

\begin{document}

\title{Dark matter scattering off ${}^4$He in chiral effective field theory  
}


\author{J. de Vries\thanksref{e1,addr1,addr2}
         \and
         C. K{\"o}rber\thanksref{e2,addr3} 
         \and 
         A. Nogga\thanksref{e3,addr4}
         \and 
         S. Shain\thanksref{e4,addr1,addr2} 
}

\thankstext{e1}{e-mail:jordydv@nikhef.nl}
\thankstext{e2}{e-mail:physics@ckoerber.com}
\thankstext{e3}{e-mail:a.nogga@fz-juelich.de}
\thankstext{e4}{e-mail:sachinshain@gmail.com}


\institute{Institute for Theoretical Physics Amsterdam and Delta Institute for Theoretical Physics, University of Amsterdam, Science Park 904, 1098 XH Amsterdam, The Netherlands \label{addr1}
           \and
           Nikhef, Theory Group, Science Park 105, 1098 XG, Amsterdam, The Netherlands \label{addr2}
           \and
           Institut f\"ur Theoretische Physik II, Ruhr-Universit\"at Bochum,  D-44780 Bochum, Germany \label{addr3}
           \and 
           Institute for Advanced Simulation (IAS-4) and Center for Advanced Simulation and Analytics (CASA), Forschungszentrum J\"ulich, D-52425 J\"ulich, Germany \label{addr4}
}

\date{Received: date / Accepted: date}

\maketitle

\begin{abstract}
We study dark matter scattering off ${}^4$He and other light nuclei using chiral effective field theory. We consider scalar DM interactions and include both one- and two-nucleon scattering processes. 
The DM interactions and nuclear wave functions are obtained from chiral effective field theory and we work up to fourth order in the chiral expansion for the latter to investigate the chiral convergence. The results for the scattering rates can be used to determine the sensitivity of planned experiments to detect relatively light dark matter particles using ${}^4$He. We find that next-to-leading-order scalar currents are smaller than expected from power counting for scattering off ${}^4$He confirming earlier work. However, the results for two-nucleon corrections exhibit a linear regulator dependence indicating potential problems in the applied power counting. We observe a linear correlation between the, in principle not observable, D-wave probability of various light nuclei and the scalar two-nucleon matrix elements, again pointing towards potentially missing contributions. 
\keywords{dark matter \and few-nucleon systems \and effective field theory}
\end{abstract}

\section{Introduction}\label{introduction}

While no direct evidence for dark matter (DM) has been found in a laboratory experiment, there exists strong observational evidence for its existence ~\cite{Bertone:2016nfn}. 
Next-gene\-ration direct detection experiments aim to probe uncharted parameter space for a wide range of DM masses, going beyond the `traditional' WIMP (Weakly Interacting Massive Particles) regime of GeV-to-TeV DM masses~\cite{Battaglieri:2017aum,Hertel:2018aal, Schumann:2019eaa}. 
To interpret these direct detection searches in terms of the underlying DM models and to connect to cosmological aspects of DM, such as the relic density, it is important to have well-controlled theoretical 
predictions for WIMP cross-sections off atomic nuclei. These predictions are complicated due to the presence of widely separated energy scales associated with  particle, hadronic, and nuclear processes. The last decade has seen the development of effective field theory (EFT) approaches to overcome this difficulty \cite{Fan:2010gt, Fitzpatrick:2012ix,Cirigliano:2012pq,Klos:2013rwa,Hoferichter:2015ipa, Brod:2017bsw,Richardson:2021liq}. For example, by assuming that DM fields are singlets under the Standard Model (SM) gauge symmetries, the DM interactions with SM fields can be captured by a series of effective operators which are dominated by the operators of the lowest dimension. After renormalization-group evolution to lower energies \cite{DEramo:2016gos, Bishara:2018vix}, these interactions can be matched to effective interactions between DM and hadrons and nuclei, which, in turn, can be used to compute DM-nucleus scattering rates. 

 In this work, we apply the framework of chiral EFT, the low-energy EFT of QCD, to perform the last steps of this program. We consider scalar-mediated DM-SM interactions, which we match onto chiral EFT hadronic interactions. The chiral EFT power counting, often call\-ed Weinberg counting, predicts that DM-nucleon interactions (or currents) provide the leading-order (LO) contribution, whereas two-nucleon currents appear at next-to-leading-order (NLO)~\cite{Prezeau:2003sv,Cirigliano:2012pq,Klos:2013rwa,Hoferichter:2015ipa,Korber:2017ery}. This is potentially interesting as these NLO two-body currents are predicted to be relatively large and could potentially be used to unravel the underlying DM interactions from DM scattering off various nuclear isotopes. 
 
 Unfortunately, explicit computations of the two-body corrections are found to be inconclusive. While shell-model computations for ${}^{132}$Xe found $\mathcal O(10\%$-$20\%)$ corrections~\cite{Hoferichter:2016nvd}, of the expected size, calculations on lighter nuclei found smaller effects \cite{Beane:2013kca,Korber:2017ery,Andreoli:2018etf}. More problematic is that two-body corrections in DM scattering off the deuteron, ${}^3$H, and ${}^3$He, strongly depend on the details of the applied wave function and thus on the applied nucleon-nucleon potential used to generate the wave functions \cite{Korber:2017ery}. Similar conclusions were drawn in Ref.~\cite{Andreoli:2018etf}, which applied phenomenological wave functions of various light nuclei and observed a large dependence on the applied regulator used in intermediate steps of the quantum Monte Carlo calculations. These results are worrisome as they indicate a potential problem in the chiral EFT power counting for two-nucleon currents and potentially jeopardize the interpretation of WIMP-nucleus scattering. Similar  problems were recently identified in other nuclear probes of beyond-the-Standard Model physics, such as neutrinoless double beta decay \cite{Cirigliano:2018hja} and indicate the presence of additional short-range two-nucleon currents. 
 
 Our main purpose here is to extend the calculations of Ref.~\cite{Korber:2017ery} to scattering off ${}^4$He nuclei for two reasons. First of all, experiments have been proposed using a liquid $^4$He detector~\cite{Hertel:2018aal,SPICEHeRALD:2021jba}. The main motivation to use relatively light target nuclei is to get sensitivity to lighter WIMPs to which more conventional experiments, for instance those involving Xe nuclei, have less sensitivity.  Second, compared to $A=2$ and $A=3$ nuclei, the binding energy per nucleon is much larger and comparable to heavier isotopes. We are particularly interested in determining whether the conclusions of \cite{Korber:2017ery} regarding two-nucleon currents are related to the diluteness of the deuteron and ${}^3$H and ${}^3$He nuclei.

In Section~\ref{scalar_int}, our computational framework is described. Specifically, we introduce the relation of the DM scattering cross sections to a set of response functions which are calculated in following based on chiral nuclear wave functions. The calculation of the wave functions is summarized in  Section~\ref{nulcear_wave_fun}. In this section, we also discuss our Bayesian approach to quantify uncertainties due to the limiting the wave function calculations to finite order. Finally, we briefly introduce nuclear transition densities that connect the wave 
function to the application to DM currents. 
Our results are given in Section~\ref{discussions}. Special attention is given here to the dependence on the regulator applied in the calculations. We conclude in Section~\ref{conclusions}.
Details of the definition of the densities and the partial wave decomposition of the current are deferred to the appendix.

\section{Computational framework}\label{scalar_int}
\begin{figure}[t]
	\centering
	\begin{subfigure}[b]{0.3\textwidth}
		\includegraphics[width=\textwidth]{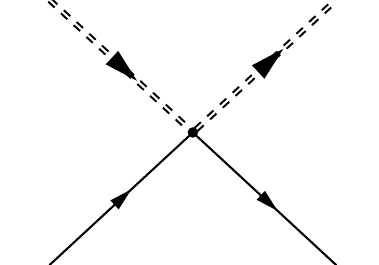}
		\caption{\label{fig:contact}}
	\end{subfigure}
	\begin{subfigure}[b]{0.3\textwidth}
		\includegraphics[width=\textwidth]{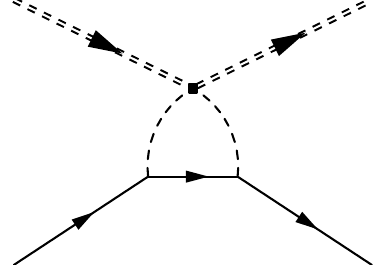}
			\caption{\label{fig:1n_pion}}
	\end{subfigure}
	\begin{subfigure}[b]{0.3\textwidth}
		\includegraphics[width=\textwidth]{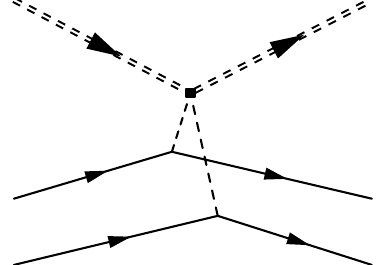}
			\caption{\label{fig:2n_2pion}}
	\end{subfigure}
	\caption{\label{fig:isoscalar_diag}Diagrams contribution to DM-nucleus scattering.  Solid lines correspond to nucleons, single dashed lines to pions and double-dashed lines to DM.  }
\end{figure}

 In this work, we follow up the computations of Ref.~\cite{Korber:2017ery} and perform first-principle computations of DM scattering off ${}^4$He isotopes. We apply chiral EFT to both generate the ${}^4$He wave functions as well as the DM currents and systematically compute the resulting DM-${}^4$He scattering rate. We select ${}^4$He because of two reasons as mentioned above: it is more sensitive for light WIMP searches, see e.g. \cite{Liao:2021npo, SPICEHeRALD:2021jba} and  the binding energy per nucleon is much larger than for $A<4$ nuclei and comparable to heavier isotopes. We are particularly interested to determine whether the conclusions of ~\cite{Korber:2017ery} regarding two-nucleon currents are related to specific spin-isospin properties and/or the diluteness of the deuteron and ${}^3$H and ${}^3$He nuclei. Much of the calculation framework follows that of Ref.~\cite{Korber:2017ery} (see also Ref.~\cite{Andreoli:2018etf}  which uses a very similar setup) and here we only give the main ingredients. Our starting point are scalar DM-SM interactions of the form 
 \begin{eqnarray}\label{Eq1}
\mathcal L_{\chi} & = & \bar \chi \chi \left (  c_u\,m_u\,\bar u u + c_d\,m_d\,\bar d d+ c_s\,m_s\,\bar s s \right. \nonumber \\
& & \quad \left. + c_G \,\alpha_s G_{\mu\nu}^a G^{\mu\nu\,a} \right )\ ,
\end{eqnarray}
where $\chi$ denotes a spin-$\tfrac{1}{2}$ DM fermion (for other DM spins, the computations are almost identical). $u$, $d$, and $s$ denote, respectively, quark fields with quark masses $m_{u,d,s}$, and $G_{\mu\nu}^a$ is the gluon field strength. We factored out one power of $\alpha_s = g_s^2/(4\pi)$. $c_{u,d,s,G}$ describe three unknown coupling constants of mass dimension $(-3)$ that parametrize the couplings strengths of DM with quarks and gluons. They can be computed in specific DM models, for example in Higgs portal models \cite{Arcadi:2019lka}. We consider the Lagrangian in Eq.~\eqref{Eq1} to be valid at relatively low energies ($\mu =1$ GeV) where we match to hadronic DM interactions. Other interactions beyond those in Eq.~\eqref{Eq1} are certainly possible, but tend to lead to suppressed two-body currents. See for instance Ref.~\cite{Hu:2021awl} for recent first-principle computations for spin-dependent cross sections. 

By application of chiral perturbation theory, the interactions in Eq.~\eqref{Eq1} can be matched to interactions between DM and nucleons, pions, and heavier hadrons. The leading-order (LO) currents involve a single nucleon and are conveniently written as
\begin{eqnarray}\label{NLO1body}
J^{(\mathrm{one-body})} (\vec q) &=&  \left[\sigma_{\pi N} - \frac{9 g_A^2 \pi m_\pi^3}{4 (4\pi f_\pi)^2}F\left(\frac{|\vec q|}{2m_\pi}\right)\right]   \,\bar c_{q^{\mathrm{(is)}}} \nonumber \\
& & -\frac{\delta m_N}{4}\bar c_{q^{\mathrm{(iv)}}}\,\tau^3_i \nonumber\\
&& +  c_s \sigma_{s}   - c_G\frac{8 \pi}{9} m_N^G\,.
\end{eqnarray}
Here we have introduced the isoscalar and isovector combinations of couplings
\begin{eqnarray}
& & \bar c_{q^{\mathrm{(is)}}}  = \frac{1}{2} \left[ c_u (1-\varepsilon) + c_d (1+\varepsilon ) \right] \,, \nonumber \\
& & \bar c_{q^{\mathrm{(iv)}}}  =   \left[c_u (1-\frac{1}{\varepsilon}) + c_d (1+\frac{1}{\varepsilon}) \right]\,,
\end{eqnarray} 
in terms of  $\varepsilon = (m_d - m_u)/(m_d + m_u) =0.36\pm0.03$\cite{ParticleDataGroup:2022pth}, and the loop function associated to Diagram \ref{fig:isoscalar_diag}(b) 
\begin{equation}
F(x) = \frac{-x + (1+2x^2)\arctan x}{3x }\,,
\end{equation}
and $F(x\ll1)\simeq 5 x^2/9 + \dots$. The various low-energy constants are given by \cite{Brantley:2016our} 
\begin{equation}
\sigma_{\pi N} = (59.1 \pm 3.5)\,\mathrm{MeV}\ ,\qquad \delta m_N = ( 2.32\pm0.17)\, \mathrm{MeV}\, ,
\end{equation}
where we used a Roy-Steiner extraction of the pion nucleon sigma term \cite{Hoferichter:2015dsa}. Lattice QCD tends to predict somewhat smaller values but might be plagued by excited-state contamination \cite{Gupta:2021ahb} and must be corrected for isospin breaking \cite{Hoferichter:2023ptl}. For the strange matrix elements we use~\cite{Borsanyi:2020bpd}
\begin{equation}
\sigma_s = m_s \frac{d m_N}{d m_s} = (37\pm 3) \MeV\,.
\end{equation}
We defined $m_N^G \equiv m_N -\sigma_{\pi N}-\sigma_s$ as the gluonic contribution to the nucleon mass.

At next-to-leading-order (NLO) Diagram \ref{fig:isoscalar_diag}(c) induces a two-body current given by
 \begin{eqnarray}\label{NLO2body}
& & J^{\mathrm{two-body}} (\vec q) \nonumber \\
& & =-m_\pi^2 \left(\frac{g_A}{2f_\pi}\right)^2 \frac{(\vec \sigma_1 \cdot \vec q_1)(\vec \sigma_2 \cdot \vec q_2)}{(\vec q_1^{\,2}+m_\pi^2)(\vec q_2^{\,2}+m_\pi^2)} \tau_1 \cdot \tau_2\,\bar c_{q^{\mathrm{(is)}}}\ ,
\end{eqnarray}
where $\vec  q_i = \vec p^{\,\prime}_i - \vec p_i$ is the difference between the outgoing and incoming momentum of nucleon $i$, and $\sigma_i$ ($\tau_i$) the spin (isospin) of nucleon $i$.

At next order in the expansion (N$^2$LO), several new diagrams appears, but they can all be absorbed into the LO one-body currents and are effectively absorbed in the values of the low-energy constants used in Eq.~\eqref{NLO1body}. At N$^3$LO there appear new two-nucleon currents both from tree- and loop-level diagrams~\cite{Krebs:2020plh} but also two-nucleon contact interactions with presently undetermined low-energy constants. One outstanding issue is how to consistently regularize the loop diagrams entering the current and the loop diagrams arising from iterating the nucleon-nucleon potential that lead to the nuclear wave functions \cite{Krebs:2019aka}. For these reasons we stick to the scalar currents up to N$^2$LO which only involve tree-level diagrams and no unknown low-energy constants. 

\subsection{Scattering cross section}
We investigate scattering processes 
$ \chi( \vec p_\chi ) + T( \vec p_T ) \rightarrow \chi( \vec p_\chi' ) + T( \vec p_T' )$, 
where $T$ denotes the target nucleus with mass $m_T$ consisting of $A$ nucleons.  The elastic unpolarized differential cross section is given by
\begin{eqnarray}
    \label{eq:def-cross_section}
& & 	\frac{\diff{\sigma}}{\diff{\vec q^2}} =
	\frac{1}{4 \pi v_\chi^2} 
	\frac{1}{2j +1} \nonumber \\ 
 & &  \qquad \quad \sum_{m_j, m_j'= -j}^j \bigg| \left\langle \Psi_T, j m_j' | \ \hat{ J }(\vec q ^2) \ | \Psi_T, j m_j \right \rangle \bigg |^2
			\ ,
\end{eqnarray}
where $\vec q $ is the momentum  transfer from DM to the target nucleus, and $v_\chi$ the DM velocity.  The wave function of the target nucleus $\ket{\Psi_T, j m_j}$  corresponds to a nucleus with total angular momentum $j$ and polarization $m_j$. 

To discuss the various one- and two-body nuclear response functions it is convenient to factor out the isoscalar piece. We follow Ref.~\cite{Korber:2017ery} and classify the cross section in terms of response functions, $\mathcal{F}^{(\nu)}_{i,\,a}(\vec q^{\,2})$, that depend on \begin{itemize}
	\item $i= \{\mathrm{is},\, \mathrm{iv},\, s,\, G\}$ which label the dependence on the DM-quark and DM-gluon couplings.
	\item The chiral order $\nu = 0,\ 1,\dots$ of the current where the dominant current for every DM interaction \\ ($\{\bar c_{q^{\mathrm{(is)}}},\, \bar c_{q^{\mathrm{(iv)}}},\, 		c_s,\, c_G \}$) begins at order $0$.
	\item We divide the NLO contributions into two-nucleon, $a=2b$, and radius corrections, $a= r$. 
\end{itemize}
We then decompose 
\begin{align}
	\frac{\diff{\sigma}}{\diff{\vec q^2}}
	=
	\bar c_{q^{\mathrm{(is)}}}^2 \frac{ \sigma_{\pi N}^2 A^2}{4 \pi v_\chi^2 }
	\bigg|
		 &\left(
			\mathcal{F}^{\left(0\right)}_{\mathrm{is}}{\left(\vec q^2\right)} + 
			\mathcal{F}^{\left(1\right)}_{\mathrm{is},\,2b}\left(\vec q^2\right)  \right. \nonumber \\ &+ \left. 
			\mathcal{F}^{\left(1\right)}_{\mathrm{is},\,r}\left(\vec q^2\right) + 
			\dots 
		\right) \nonumber \\  &+
		\alpha_{\mathrm{iv}} \left( 
			\mathcal{F}^{\left(0\right)}_{\mathrm{iv}}{\left(\vec q^2\right)}  + 
			\dots 
		\right) \nonumber \\  &+
		\alpha_{s} \left( 
			\mathcal{F}^{\left(0\right)}_{s}{\left(\vec q^2\right)}    + 
			\dots 
		\right)\nonumber \\ &+
		\alpha_{G} \left( 
			\mathcal{F}^{\left(0\right)}_{G}{\left(\vec q^2\right)}  + 
			\dots 
		\right)
		\bigg|^2\, ,\label{eq:def-response_functions}
		\end{align} 
where we kept terms up to NLO in the chiral power counting, and 
\begin{eqnarray}\label{eq:alpha}
	& & \alpha_{\mathrm{iv}} = -\left(\frac{\delta m_N}{4 \sigma_{\pi N}}\right) \frac{ \bar c_{q^{\mathrm{(iv)}}}}{\bar c_{q^{\mathrm{(is)}}}}, \qquad 
	\alpha_s =  \left(\frac{\sigma_{s}}{ \sigma_{\pi N}}\right)\frac{c_s}{\bar c_{q^{\mathrm{(is)}}}}, \nonumber \\  	
	& & \alpha_G= \left(- \frac{8 \pi}{9} \frac{m^G_N}{\sigma_{\pi N}}\right) \frac{c_G}{\bar c_{q^{\mathrm{(is)}}}}\ .
\end{eqnarray}
Many response functions are actually equal up to this order in the power counting and we have
\begin{eqnarray}\label{normalization}
	\mathcal{F}^{\left(0\right)}_{\mathrm{is}}(\vec q^2)
	&=&
	\mathcal{F}^{\left(0\right)}_{s}(\vec q^2)
	\,\,=\,\,
	\mathcal{F}^{\left(0\right)}_{G}(\vec q^2)\, ,
\end{eqnarray}
which are normalized to $\mathcal{F}^{\left(0\right)}_{\mathrm{is}}(0)=1$. The radius correction $\mathcal{F}^{\left(1\right)}_{\mathrm{is},\ r}\left(\vec q^2\right)$ also involves the same nuclear information as $\mathcal{F}^{\left(0\right)}_{\mathrm{is}}$ except for an additional overall dependence on $\vec q^2$. We therefore need to perform nuclear calculations of $\mathcal{F}^{\left(0\right)}_{\mathrm{is}}(\vec q^2)$, $\mathcal{F}^{\left(0\right)}_{\mathrm{iv}}(\vec q^2)$, and $\mathcal{F}^{\left(1\right)}_{\mathrm{is},\ 2b}\left(\vec q^2\right)$, which we will present below for various target nuclei.

\section{Nuclear wave function and scattering matrix elements}\label{nulcear_wave_fun}

\subsection{Nuclear wave functions}
We use the momentum-space basis to calculate the wave function and matrix elements involving the DM interactions discussed above. ${}^2$H, ${}^3$He, and  ${}^4$He wave functions are evaluated by solving the non-relativistic Schr\"{o}dinger equation in momentum-space. Ref.~\cite{Korber:2017ery} and Ref.~\cite{Nogga:2001cz} shows detailed calculation of ${}^2$H and ${}^3$He, and ${}^4$He  wave functions, respectively. In this section, we summarize the relevant parts of the wave function calculations. 

The wave function of deuteron ($d$) is easily evaluated by solving
\begin{align}
    \ket{\psi_d} = \frac{1}{E_d-T} V_{12} \ket{\psi_d}\,,
\end{align}
where $E_d$ is the deuteron binding energy, $T$ is the two-nucleon ($N\!N$) kinetic energy, and $V_{12}$ is
the $N\!N$ potential. It is convenient to use the partial wave basis ($\ket{p\alpha}$) for the wave function, where $p$ is the momentum of the deuteron and $\alpha$ represents the partial wave contributing to the deuteron bound state. The deuteron bound state has a total orbital angular momentum $l_{12} = 0,2$, a spin $s_{12}=1$, and a total angular momentum $j_{12}=1$. 

We use modern phenomenological $N\!N$ interactions and \chipt \ to calculate the nuclear wave functions. We calculate the former by using the standard AV18~\cite{Wiringa:1994wb} and
CD-Bonn~\cite{Machleidt:2000ge} phenomenological interactions. Since these two interactions have different non-observable kinetic energies, potential energies, and D-state probabilities, they provide an added benefit of checking model dependence~\cite{Korber:2017ery}.

We mainly use semilocal coordinate-space regularized (SCS) chiral interactions from Ref.~\cite{Epelbaum:2014sza} to calculate the chiral wave functions up to \nlo{4} chiral order.  For ${}^4$He, we also show some results based on the semilocal momentum-space (SMS) regularized interactions from Ref.~\cite{Reinert:2017usi} which include \nlo{5} contact interactions in the F-waves, these are labeled \nlo{4}${}^+$. 
The approach successfully describes various $N\!N$ observables to a high degree of accuracy provided that the interactions are regulated using finite value cut-offs. The SCS interactions use a regularization defined in the configuration space and parameterized by a short-distance scale $R$. In Ref.~\cite{Epelbaum:2014efa}, an optimal range of cut-offs $R$ between $0.8\fm$ and $1.2 \fm$ has been identified. For this range of cutoffs, the LECs are of natural size and no spurious bound states exists. It was also found that the best description of $N\!N$ data can be achieved for $R=0.9 \fm$. Since an electromagnetic (EM) interaction accompanies AV18, we included the same EM interactions for the CD-Bonn and the chiral potentials except for the neutron-proton system since the relevant parts are here already included in the strong part. 

The calculation of the $^3$He and $^3$H wave functions was summarized in Ref.~\cite{Korber:2017ery}. 
The wave functions for ${}^4$He ($\ket{\Psi}$) are obtained by rewriting the Schr\"{o}dinger equation into a set of two Yakobovsky equations~\cite{Nogga:2001cz}
\begin{align}
    \ket{\psi_{1A}} &\equiv \ket{\psi_{(12)3,4}} =  G_0 t_{12} P[(1-P_{34}) \ket{\psi_{1A}} + \ket{\psi_{2A}}] \nonumber \\
    & \qquad + (1 + G_0t_{12} )V^{(3)}_{123} \ket{\Psi}\,,
    \\
    \ket{\psi_{2A}} &\equiv \ket{\psi_{(12)34}} =  G_0 t_{12} \tilde{P}[(1-P_{34}) \ket{\psi_{1A}} + \ket{\psi_{2A}}] ,
\end{align}
where we have introduced two Yakubovsky components $\ket{\psi_{1A}}$ and $\ket{\psi_{2A}}$. The $N\!N$ interactions enter into this equation by the $N\!N$ T-matrix $t_{12}$ and $G_0$ denotes the free 4N propagator. From the three-nucleon force (3NF) the part $V_{123}^{(3)}$ symmetric in subsystem (12)  is applied. The total 3NF is obtained by cyclic permutations as 
\begin{align}
    V_{123} = V_{123}^{(1)} + V_{123}^{(2)} + V_{123}^{(3)}.
\end{align}
In our previous study, we found that the contribution of the 3NFs to the matrix elements relevant for dark matter scattering is minor \cite{Korber:2017ery}. Therefore, we will neglect this part in the calculations below. 
Using the antisymmetry of the four-nucleon wave functions, it is possible to relate different Yakubovsky components using the permutation operators $P=P_{12}P_{23} + P_{13} P_{23},  P_{34}$ and  $\tilde P = P_{14}P_{23}$. The total 4N wave function can then be obtained applying these permutation operators to the Yakubovsky components 
\begin{align}
    \ket{\Psi} & = [1-(1+P)P_{34}](1+P)\ket{\psi_{1A}} \nonumber \\
    & \quad + ( 1+P)(1+ \tilde{P})\ket{\psi_{2A}}
\end{align}
The advantage of using Yakubovsky components for these bound state calculations is that the partial wave decomposition converges faster for the Yakubovsky components than for the wave function and also because $\psi_1$ and $\psi_2$ are expanded independently in 3+1 and 2+2 Jacobi momenta, respectively. 

Fig.~\ref{fig:jacobi} of the appendix defines these Jacobi coordinates. Explicitly, the basis states 
for 3+1 coordinates are given by 
\begin{align}
   \label{eq:psi31}
    & \ket{p_{12}p_3 p_4 \alpha} \nonumber \\ & = 
    \left| p_{12}p_3 p_4 \left[ \left( (l_{12}s_{12})j_{12} (l_{3}\,   1/2 )I_3 \right) j_3 (l_{4} \, 1/2 )I_4 \right] J \right. \nonumber \\
    & \phantom{.}\qquad \qquad \qquad \left. \left( (t_{12}\,  1/2)\tau_3 1/2 \right) T   \right\rangle 
\end{align}
where $p_{12}, p_3$ and $q_4$ are magnitudes of the Jacobi relative momenta between particle (12), the third particle and the (12) subsystem and the fourth particle and the (123) subsystem, respectively. The orbital angular momenta related to these momenta are denoted by $l_{12}, l_3$, and $l_4$ and are coupled to the spin of the (12) system $s_{12}$, and the spin-$\tfrac{1}{2}$ of particles 3 and 4  to intermediate angular momenta $j_{12}, I_3$, and $I_4$. We indicate the coupling scheme using the brackets in Eqs.~(\ref{eq:psi31})  and (\ref{eq:psi22}). $j_{12}$ and $I_3$ are coupled to the total angular momentum of the (123) system, $j_3$, which is when coupled with $I_4$ to the total angular %
momentum of the ${}^4$He state $J=0$. The isospin of the (12) subsystem $t_{12}$ is coupled with the isospin of the third particle to $\tau_3$ and with the fourth particle to the total isospin of ${}^4$He $T=0$. 

For 2+2 coordinates the basis states are 
\begin{align}
    \label{eq:psi22}
    & \ket{p_{12} p_{34} q \beta} = \left|  p_{12}p_{34} q  \left( \left[ (l_{12}s_{12})j_{12} \lambda \right]  I \, (l_{34}s_{34})j_{34} \right) J \right. \nonumber \\ 
    & \qquad \qquad \qquad \left.   (t_{12}\,  t_{34}) T \right\rangle 
\end{align}
where $p_{12}, p_{34}$, and $q$ are magnitudes of the Jacobi relative momenta between particle (12) and (34)  and relative momentum of  (12) and (34) subsystems, respectively. The additional orbital angular momenta related to these momenta are denoted by $l_{34}$ and $\lambda$. $l_{12}$ and $l_{34}$  are coupled to the spin of the (12) and (34) systems, $s_{12}$ and $s_{34}$, to subsystem angular momenta $j_{12}$ and $j_{34}$.  $j_{12}$ and $\lambda$ are coupled to the intermediate  angular momentum $I$ which is then coupled with $j_{34}$ to the total angular momentum  $J=0$. The isospin of the (12)  and (34) subsystems are coupled to the total isospin $T=0$.  

For our calculation, we restrict the orbital angular momentum to 6, $j_{12}$ to 5 and the sum of the orbital angular moment $l_{12}+l_3+l_4$ and $l_{12}+l_{34}+\lambda$ to 10. With this restriction, we achieve binding energy converged up to 20 keV. More details can be found in Ref.~\cite{Nogga:2001cz}.

\subsection{Chiral expansion and uncertainty estimation} \label{uncertainity}

On top of the chiral expansion scheme associated with the dark matter current operators, also the wave functions are obtained by chiral power counting scheme.
The expansion of the wave functions follows chiral effective field theory\cite{Reinert:2017usi} for describing the target nucleus in the absence of dark matter.
The parameter for this expansion is associated with a momentum scale at the order of binding momenta and the pion mass $\sim Q_\psi \sim m_\pi/\Lambda_\chi \sim \gamma/\Lambda_\chi$, where $\Lambda_\chi \sim 1$ GeV, is the chiral EFT breakdown scale and $\gamma$ denotes the binding momentum.
The current expansion considers an external scalar source, dark matter \cite{Krebs:2020pii}, which also has an expansion in $Q_\psi$ but also obtains corrections scaling as $ Q_J \sim q/\Lambda_\chi$, where $q$ denotes the  momentum transfer. For the currents we will collectively denote the scaling with $Q_J$ to also account for corrections scaling as $m_\pi/\Lambda_\chi \sim \gamma/\Lambda_\chi$.

The truncation error of matrix elements are dominated by the largest corrections to either source.
That is, performing the computation at a fixed order of the wave functions $\nu$ and the current operator $\mu$, the matrix element is expected to scale as
\begin{align}
    & \ket{\psi^{(\nu)}} \equiv \ket{\psi} [1 + \mathcal O(Q_{\psi}^{\nu + 1})] \,, \\
    & \hat{J}^{(\mu)} \equiv \hat{J} [1 + \mathcal O(Q_{J}^{\mu + 1})] \,, \\
    & \Rightarrow \braket{\psi^{(\nu)} | \hat{J}^{(\mu)} | \psi^{(\nu)}}  \nonumber \\
    & \qquad = \braket{\psi |  \hat{J} | \psi }[ 1 + \max(O(Q_{J}^{\mu + 1}), O(Q_{\psi}^{\nu + 1}))]\,.
\end{align}
Obviously, it depends on the orders $\mu$ and $\nu$ up to which the currents and the wave functions were expanded, respectively, and on the expansion scales $Q_J$ and $Q_\psi$ which contribution drives the uncertainty of the matrix element. 
For external currents at a momentum scale close to the nuclear scale, i.e. $Q_J \approx Q_\psi$, it makes sense to combine power countings~\cite{Filin:2020tcs} so that the maximum of both corrections can be related to the minimum of respective orders.
As such, one has to simultaneously increase both orders to make more accurate predictions.

To test the accuracy in the description of the scalar-DM matrix elements, we test the convergence behavior for DM-currents in dependence of the chiral order of the wave functions
\begin{equation}
    |\braket{\psi^{(\nu)} | \hat{J}^{(\mu_{\rm fixed})} | \psi^{(\nu)}}|^2 \equiv J^{(\nu)}(\mu_{\rm fixed}) \,,
\end{equation}
This allows us to employ standard chiral uncertainty estimation schemes\cite{Furnstahl:2015rha} for the matrix element, where the converged matrix element $O_J^{(\infty)}$ is equal to the matrix element at a finite order plus corrections
\begin{equation}
    J^{(\infty)}(\mu_{\rm fixed})
    =
    J^{(\nu)}(\mu_{\rm fixed})
    +
    \delta J^{(\nu)}(\mu_{\rm fixed})
    \,.
\end{equation}
This correction, the truncation error, follows the expansion
\begin{eqnarray}
\label{eq:cnexpansion}
    & & J^{(\nu)}(\mu_{\rm fixed}) \leftrightarrow J^{\rm ref} \sum_{n=0}^{\nu} c_n Q_{\psi}^n \,, \nonumber \\
    & & \delta J^{(\nu)}(\mu_{\rm fixed}) \leftrightarrow J^{\rm ref} \sum_{n=\nu+1}^{\infty} c_n Q_{\psi}^n \,.
\end{eqnarray}
For a fixed reference scale $J^{\rm ref}$ and expansion parameter $Q_{\psi}$, knowing the values of all expansion coefficients $c_n$ allows to infer the size of the truncation error.
In particular, we apply a Bayesian inference scheme where we relate the distribution of expansion coefficients to the final observable.
As such the posterior predictive distribution of the truncation error is obtained by integrating over (marginalizing) the posterior distribution of the expansion coefficients $P(\{ c_n \} | M, D)$ times the distribution of the observable given the coefficients (i.e., a delta distribution)
\begin{align}
    & P(\delta| \nu; M,  D) \nonumber \\
    &\sim
    \left(
    \prod_n
    \int d c_n
    \right)
    \delta (\delta - \delta J^{(\nu)} (\{ c_n \})) P(\{ c_n \} | M, D)
    \,.
\end{align}
The data points $D$ correspond to observables computed at different chiral orders $D = \{ J^{(\nu)}(\mu_{\rm fixed}) \}$, times a prior distribution $P(\{ c_n \} | M)$ for the expansion coefficients
\begin{align}
    P(\{ c_n \} | M, D)
    &\sim
    \exp\left\{
        - \chi^2(\{ c_n \}, D)/2
    \right\}
    P(\{ c_n \} | M)\,.
\end{align}
The letter $M$ indicates that your probabilities also depend on the model, i.e. the ansatz in Eq.~\eqref{eq:cnexpansion}. 
The posterior distribution is given by a data likelihood distribution ($\sim \exp\{ -\chi^2/2\}$) with
\begin{equation}
    \chi^2 = \sum_{\nu} \left( J^{\nu} - J^{\rm ref}\sum_{n=0}^{\nu} c_n Q_{\psi}^n \right)^2 \,.
\end{equation}
The $\chi^2$ distribution employed here does not accommodate any additional uncertainty as the only uncertainty for the data points $D$, in this case, are numerical errors which we assume to be uncorrelated and of equal size for each computation.
Thus they correspond to an overall normalization factor not important for this analysis.
The employed prior, following the prescription in the literature, aims at quantifying the naturalness assumption.
This estimation is enabled by the software \texttt{gsum}\cite{Melendez:2019izc}\footnote{With \texttt{df=0.6}, \texttt{scale=0.8} and the expansion parameter $Q_\psi = (m_\pi^k + q^k) / ( \Lambda_b(m_\pi ^ {k - 1} + q ^ {k - 1})) $, where $q$ is the scattering momentum, $m_\pi$ the pion mass and $\Lambda_b \sim 600$ MeV the breakdown scale. The choice of $k$ is not fixed and we follow \cite{Melendez:2019izc} and set $k=8$.}.

\subsection{Density matrix formalism}

In this section, we describe the theoretical framework for calculating matrix elements in Eq.~\eqref{eq:def-cross_section}. We can separate this calculation into two parts: the first part is the  \textit{interaction kernel}, which captures the one- and few-body currents, and the second part is the \textit{structure}, which contains the accurate eigenstates of the Hamiltonian of the nucleus. 

In the \textit{traditional method}, we calculate the cross-section by integrating the interaction kernel directly with the wave function of the nucleus.
\begin{align}
    \braket{\Psi'| O | \Psi}_{\rm traditional}
    &=
    \int \{\dd p_i\} \Psi^{'\dagger}(\{p_i\}) \Psi(\{ p_i\}) O(\{ p_i\}) \,,
\end{align}
where $\{ p_i\}$ are the momentum of the internal nucleons $i$, the quantum numbers of the nucleons are suppressed for simplicity.

Ref.~\cite{Griesshammer:2020ufp} introduced an alternative method using \textit{transition density amplitudes } $\rho$ (referred to as \textit{densities})
\begin{align}
    \braket{\Psi'| O | \Psi}_{\rm density \ formalism}
    &=
    \sum_{\{  n_i\} } \rho(\{ n_i\}) \otimes O(\{ n_i\})   \,,
\end{align}
where $\{n_i\} $ are the quantum number of the internal nucleons $i$. This approach significantly improves the computational time; for Compton scattering the computation time gets reduced by a factor of ten~\cite{Griesshammer:2020ufp}. It also separates the calculations of structure part and operator part so that it is more easy to apply the same operators to different nuclei. Finally, the densities and operators can be independently tested mitigating the danger of computational mistakes. The densities are freely available using a python package for accessing the files \cite{Nogga:2023nucdens}. 
We have used this method to calculate the matrix elements and we found an order-of-magnitude speed-up. This works because  the densities do not depend on the interaction probes (in our case, scalar dark matter interaction) and the nuclear-structure part of the calculation gets factorized from the interaction kernel part. 

For any external-probe matrix element, we need to read these nuclear densities and do convolution with the appropriate interaction kernels that encode the one- and two-body current operators on the momentum-spin basis. The computational effort associated with this nuclear structure piece of the calculations increases significantly with $A$, but highly parallelized and optimized codes exist that solve the Schr\"odinger equation and obtain  the wave functions of light nuclei. Constructing densities from those wave functions is straightforward. In short, once we generate the nuclear densities using a supercomputer, all future matrix element calculations can be performed on personal computers by reading these densities. A detailed calculation of one- and two-body densities is presented in Ref.~\cite{Griesshammer:2020ufp} for $A=3$. In \ref{Appx_density}, we briefly introduce the generalization to $^4$He.

\section{Results and Discussion}\label{discussions}
We calculated the response functions $\Fis{0,1}$ for transfer momenta in the range $q=(0-200)$\ MeV using wave functions of different chiral orders and a range of cut-offs $\{R_2,R_3,R_4,R_5\}=\{0.9,1.0,1.1,1.2\} $ fm (we have omitted $R_1 =0.8$ fm as it leads to a poorer description of the $N\!N$ data than $R_2$). Before we discuss the wave function dependence and theoretical uncertainties of our computations, we first analyze the general features using a fixed wave function obtained from the cut-off $R_2$. 

\begin{figure*}[t]
    \centering
    \begin{subfigure}[b]{.45\textwidth}
        \centering
        \includegraphics[width=\textwidth]{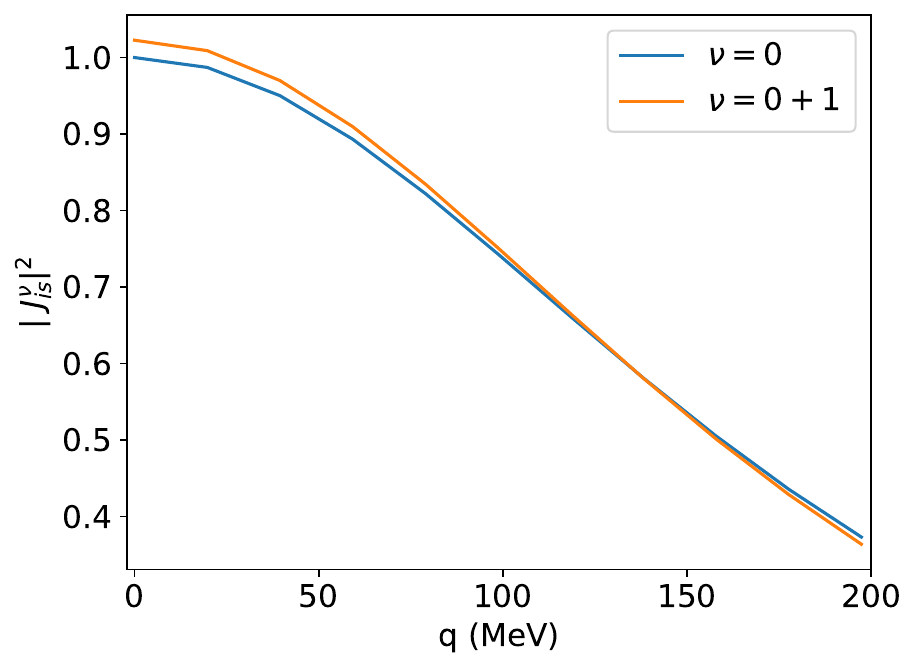}
        \caption{\label{fig:Fis_01_2H}${}^2$H}
    \end{subfigure}
    \hfill
    \begin{subfigure}[b]{.45\textwidth}
        \centering
        \includegraphics[width=\textwidth]{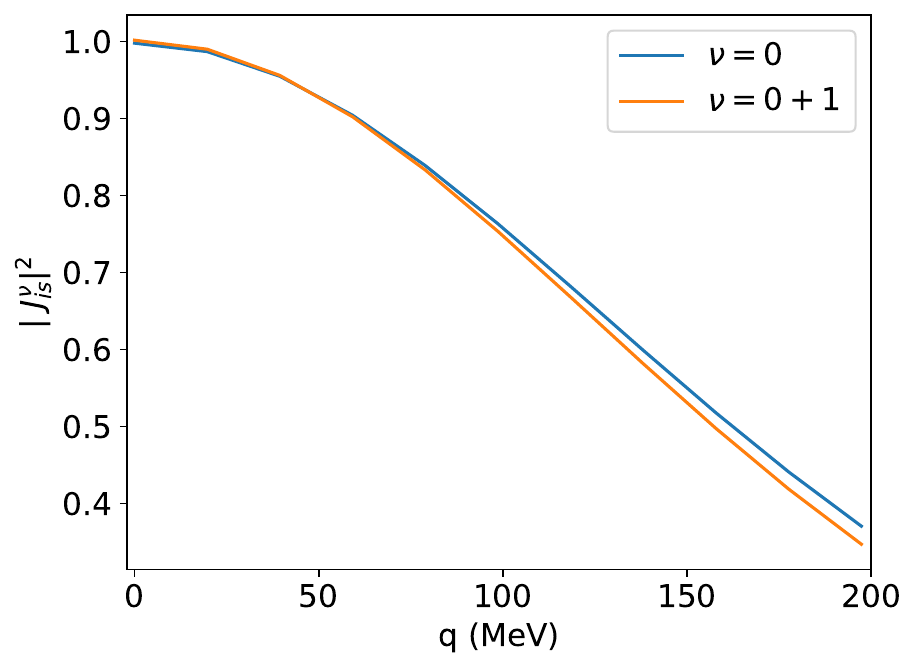}
        \caption{\label{fig:Fis_01_3H}${}^3$He}
    \end{subfigure}
    \\
    \begin{subfigure}[b]{.45\textwidth}
        \centering
        \includegraphics[width=\textwidth]{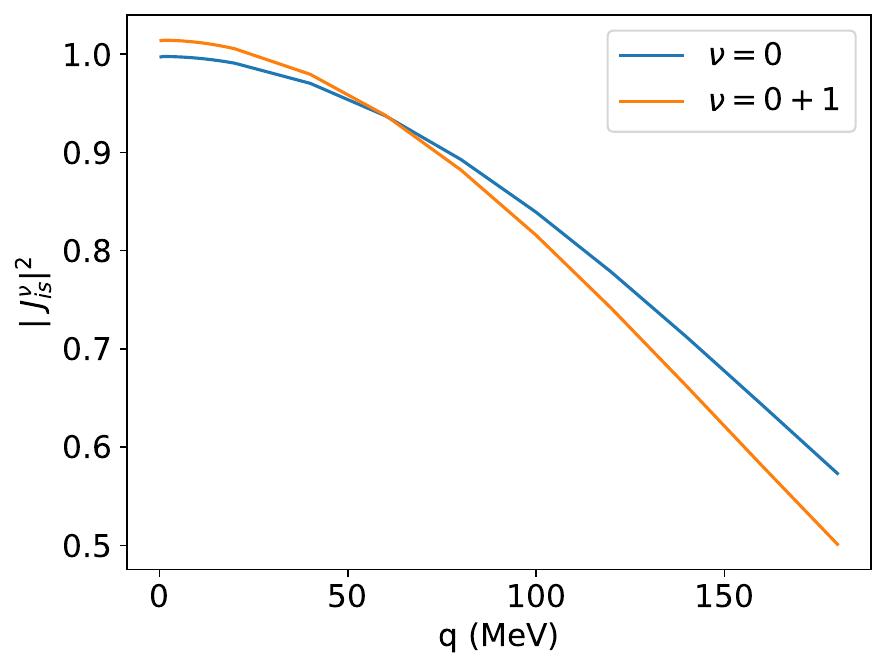}
        \caption{\label{fig:Fis_01_3He}${}^4$He}
    \end{subfigure}
    \hfill
    \caption{Isoscalar structure functions of ${}^2$H, ${}^3$He, and ${}^4$He as a function of transfer momentum $q$. We apply \nlo{4} wave functions with a fixed cut-off $R_2 = 0.9$ fm. The chiral order $\nu = 0 \ (0+1)$ is shown by the blue (orange) line. }
    \label{fig:Fis_01}
\end{figure*}

In Fig.~\ref{fig:Fis_01}, we present the response functions for  ${}^2$H, ${}^3$He, and ${}^4$He nuclei as function of the transferred momentum. The maximum momentum transfer is related to the DM and target mass and the DM velocity distribution. Using the standard distribution and the DM escape velocity $v_{\chi}^{\mathrm{esc}} \simeq 550$ km/s~\cite{Smith:2006ym}, and assuming $m_\chi \sim m_{{}^4\mathrm{He}}$ the maximum momentum transfer is roughly $|\boldq| \leq 3\times A$ MeV, so about $12$ MeV for ${}^4$He \cite{Korber:2017ery}. We show results for larger momentum transfer as well to analyze the accuracy and cut-off dependence of our results, and for non-standard DM velocity distributions. 

The LO response function $\abs{\Fis{0}}^2$, including just the one-nucleon DM interactions, is shown for the three light nuclei in Fig.~\ref{fig:Fis_01} in blue. The result use N${}^4$LO wave functions and a fixed cut-off.
 At zero momentum transfer, the LO response functions equal unity  because of the normalization condition. The NLO response function is defined as
\begin{align}
    \abs{\Fis{0+1}}^2 \equiv \abs{\Fis{0}+\Fisa{1}{r}+\Fisa{1}{2b}}^2 \, 
\end{align}
and includes the radius and two-body currents.  It is shown in orange. We conclude that $\Fis{0}$ is a good approximation for the considered range of transfer momentum. The NLO contributions only change the LO results by a few percent for ${}^2$H, ${}^3$He, and ${}^4$He. Despite the larger binding energies, the NLO corrections does not grow for ${}^4$He and remain significantly smaller than power-counting estimates. It is intriguing that much larger corrections, up to $20$-$30$\%, are found for heavier systems \cite{Hoferichter:2016nvd}, and it would be interesting to find for which size of nucleus this behavour kicks in. For ${}^2$H and ${}^3$He, the NLO corrections decrease for larger transfer momentum due to a cancellation between radius and two-body NLO corrections. This is not true for ${}^4$He where both terms have the same sign although, as discussed below, this does depend on the applied regulator.

 \begin{figure}[h!t]
     \centering
     \includegraphics[width=.5\textwidth]{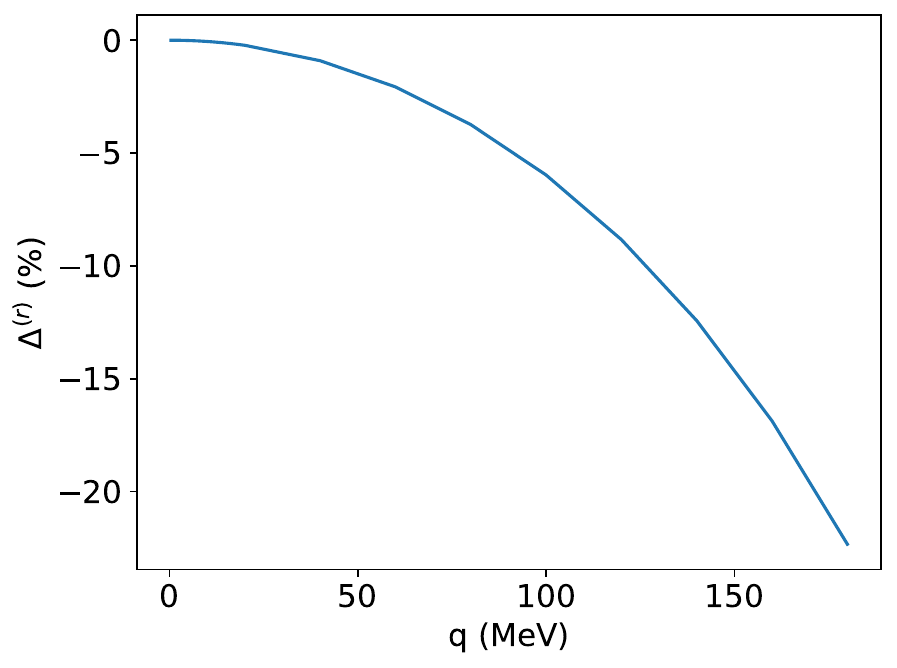}
     \caption{The percentage relative contribution of radius corrections of ${}^4$He as a function of transfer momentum $q$ for cut-off $R_2$ and the \nlo{4} wave function. }
     \label{fig:delta_r}
 \end{figure}

\begin{figure*}[h!t]
    \centering
    \includegraphics[width=\textwidth]{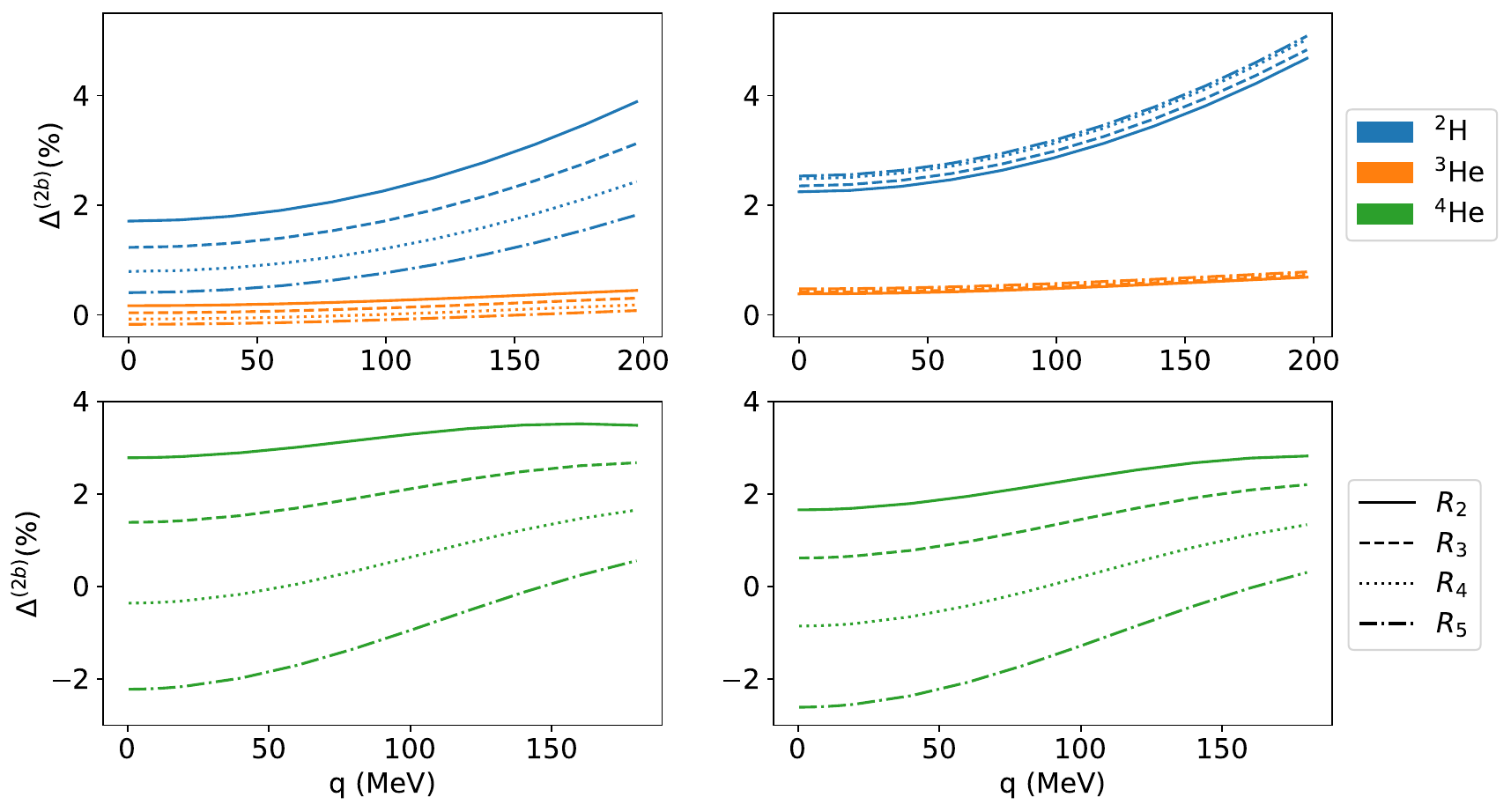}
    \caption{The percentage relative contribution of two-body corrections of ${}^2$H, ${}^3$He, and ${}^4$He as a function of transfer momentum $q$ for various cut-offs for \nlo{2} (left) and \nlo{4} (right) chiral wave functions. }
    \label{fig:delta_2b}
\end{figure*}

\begin{figure*}[h!t]
    \centering
    \includegraphics[width=\textwidth]{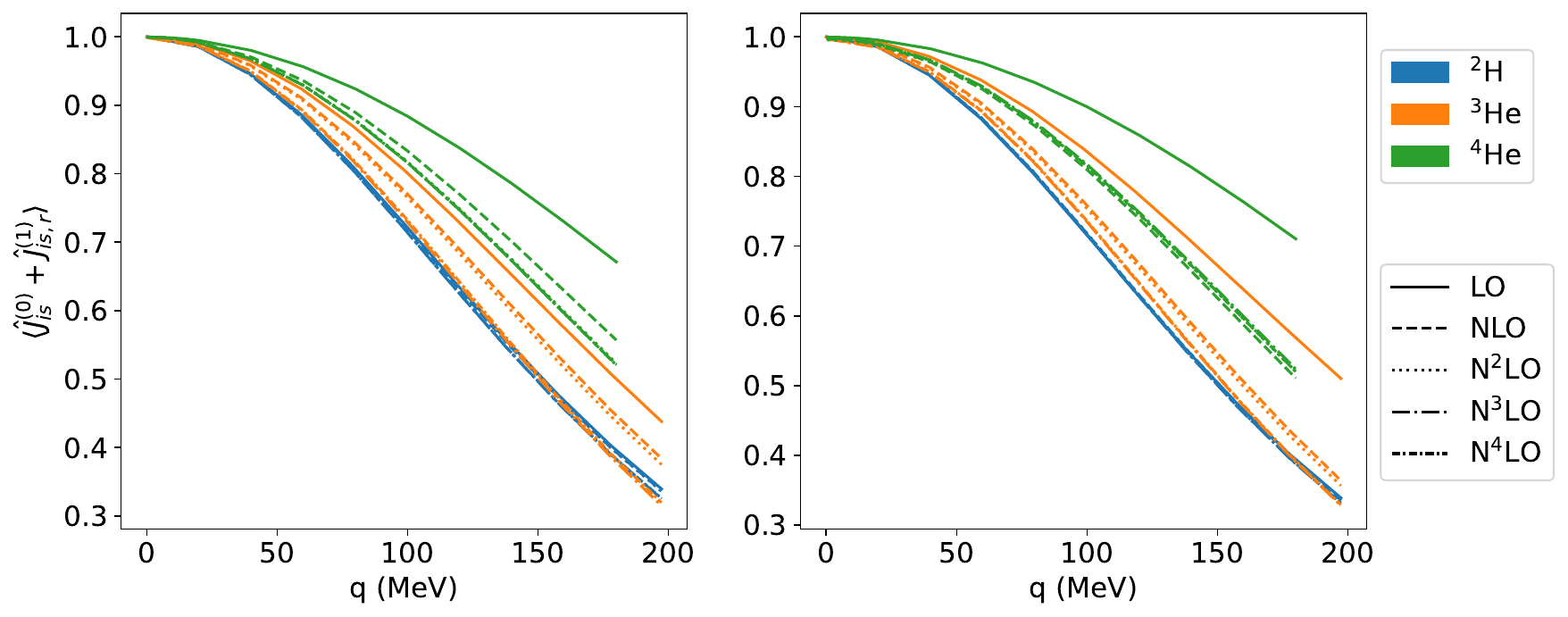}
    \caption{Isoscalar one-body structure functions as a function of transfer momentum $q$ for LO to \nlo{4} chiral wave functions. The left (right) panel matrix elements are calculated using cut-off $R_5 = 1.2 \fm$ ($R_2 = 0.9 \fm$).  }
    \label{fig:one_body}
\end{figure*}

To capture the details of the NLO contribution, we introduce the relative radius and two-body corrections
\begin{align}
    \Delta^{(r)}
    &=
    \frac{
    \abs{\Fis{0+1}}^2 - \abs{\Fis{0}+\Fisa{1}{2b}}^2
    }{\abs{\Fis{0+1}}^2}\,,
    \nn \\
    \Delta^{(2b)}
    &=
    \frac{
    \abs{\Fis{0+1}}^2 - \abs{\Fis{0}+\Fisa{1}{r}}^2
    }{\abs{\Fis{0+1}}^2}\,.
    \label{eq:rad_2b_corr}
\end{align}
The radius contribution is given by
\begin{align}
    \Delta^{(r)} & \simeq
    \frac{2 \Fisa{1}{r}}{\Fis{0+1}} \simeq
    -\frac{2}{\sigma_{\pi N}}\frac{9 g_A^2 \pi \mpi^3}{4(4\pi \Fp)^2} F\left(\frac{\abs{\boldq}}{2\mpi}\right)
    \nonumber \\
    & \xrightarrow{\abs{q}\ll 2 \mpi}
    -\frac{5 g_A^2 \pi \mpi \abs{\boldq}^2}{8 (4\pi \Fp)^2 \sigma_{\pi N}} \,.
    \label{delta_r_approx}
\end{align}
The radius contribution vanishes at lower momentum but becomes dominant at higher momentum and makes the net NLO correction negative. Furthermore, all the nuclear effects drop out and $\Delta^{(r)} $ shows identical momentum dependence for all nuclei, which has been verified by our results. In Fig.~\ref{fig:delta_r}, we show the momentum dependence of $\Delta^{(r)}$ for the ${}^4$He nucleus.  We omitted other nuclei and cutoffs because of the identical behavior. 

In Fig.~\ref{fig:delta_2b}, we present our results on $\Delta^{(2b)}$ for various nuclei. In the left panels, we apply N${}^2$LO chiral wave functions and show results for the four applied regulators. The two-body corrections are small and lie well below power-counting estimates as was found in Refs.~\cite{Korber:2017ery,Andreoli:2018etf}. What is worrisome is that the regulator dependence is large, ranging from $\Delta^{(2b)}({}^3\mathrm{He}) = (-0.01 \pm 0.13)\%$ to $\Delta^{(2b)}({}^4\mathrm{He}) = (0.4\pm 1.9)\%$ at $q^2=0$. Even the signs are uncertain. One might hope this would improve once higher-order wave functions are applied and the corresponding results are shown in the right panels. The ${}^2$H and ${}^3$He results are indeed essentially regulator independent for N${}^4$LO wave functions, confirming results found in Ref.~\cite{Korber:2017ery}. However, no such convergence is shown for ${}^4$He as clearly seen in bottom-right panel of Fig.~\ref{fig:delta_2b}. 

This is puzzling. The power counting indicates that at this order the only NLO currents are coming from one-body radius corrections and the pion-range two-body corrections. There are no free counter terms so the calculations should show regulator-independent matrix elements. We stress that similar cut-off dependence is seen in quantum Monte Carlo calculations of DM response functions for light nuclei using phenomenological $N\!N$ potentials \cite{Andreoli:2018etf}.  
In Fig.~\ref{fig:one_body}, we demonstrate that the power-counting issues only affect the two-body currents. The one-body corrections show good convergence for all nuclei even for N${}^2$LO wave functions. 

\begin{figure*}[h!t]
    \centering
    \includegraphics[width=\textwidth]{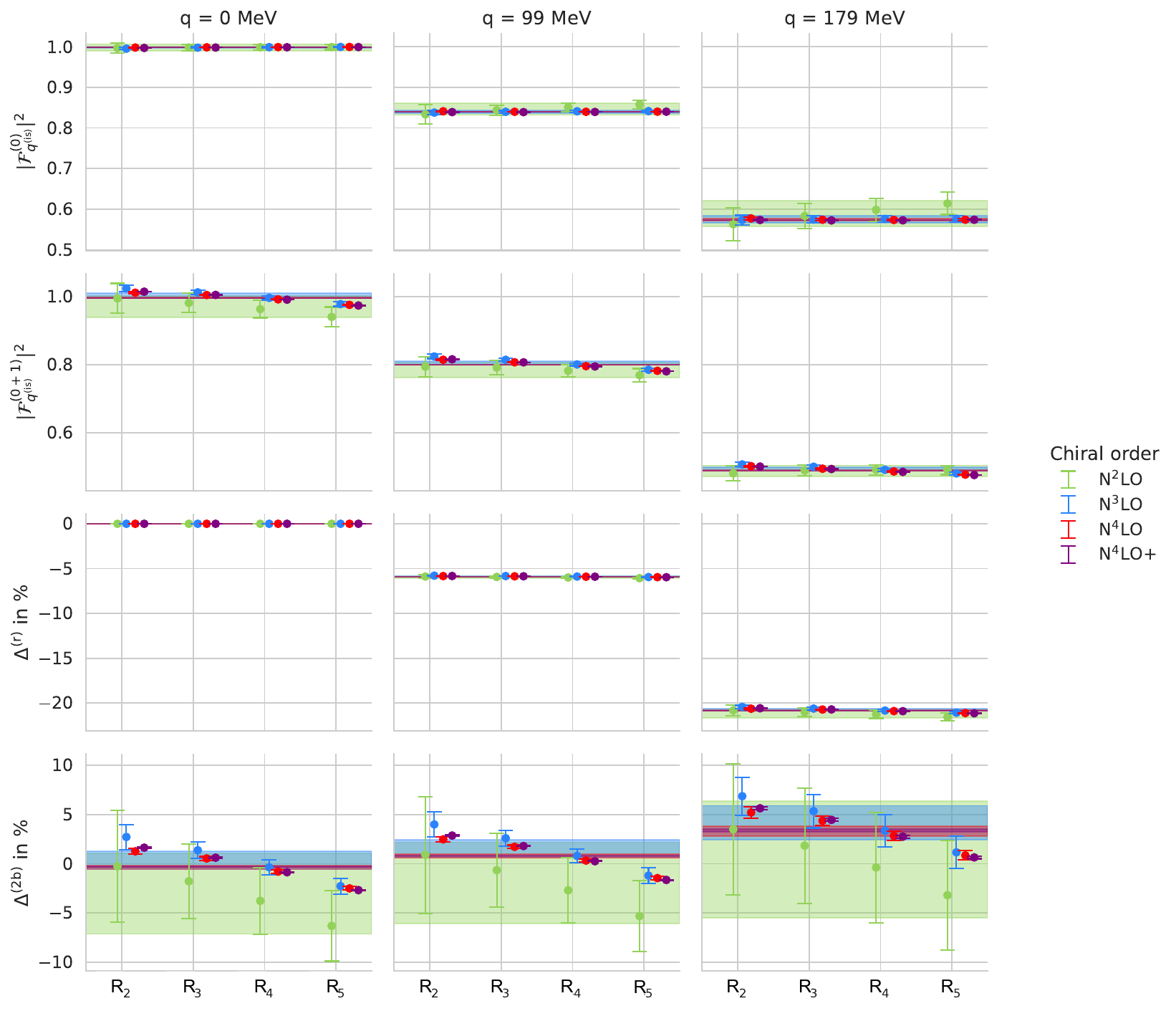}
    \caption{Isoscalar structure functions of ${}^4$He up to chiral order $\nu = 0+1$ as a function of cut-offs for  different transfer momentum $q$ . The first and second panel indicate the leading order  ($\nu=0$) one-nucleon isoscalar structure functions and the full order ($\nu=0+1$) isoscalar structure functions, respectively. The third (fourth) panel indicate the percent radius (two-body) corrections. We present the results for chiral wave functions (\nlo{2} to \nlo{4}$^+$). We omitted the results for LO and NLO wave functions due to the large uncertainty bands.}
    \label{fig:4he_lambda}
\end{figure*}

In Fig.~\ref{fig:4he_lambda}, we present our findings in a different way. We plot  LO and NLO isoscalar structure functions of ${}^4$He for three values of transfer momenta $q$ and four cut-offs. We have included different chiral order wave functions to check the convergence of the matrix elements, LO and NLO wave functions are not included. Apart from $\Fisa{1}{2b}$, the isoscalar structure functions display order-by-order convergence for wave functions and exhibit regulator independence. The $\Fis{0}$ shows the expected decrease in values for higher values of transfer momentum, which is in agreement with Ref.~\cite{Korber:2017ery}. The increase in uncertainty of $\Fis{0}$  for higher transfer momentum is the reflection of the complicated nature of four-nucleon wave functions. 

NLO contributions occur only at few percent level. The relatively minute uncertainty of $\Delta^{(r)}(\boldq^2)$ can be understood as the correction is proportional to the  $\Fis{0}$, see Eq.~\eqref{delta_r_approx}. 
The two-body corrections account for almost all the uncertainties associated with the NLO corrections. 
The bottom panel of Fig.~\ref{fig:4he_lambda} shows that $\Fisa{1}{2b}$ fails to converge to a fixed value for different regulators and for highest-order wave functions shows a linear dependence on the applied cut-off for all values of transfer momenta. 

The uncertainty bands in Fig.~\ref{fig:4he_lambda} have been obtained from the prescription in Sect.~\ref{uncertainity}. Each uncertainty band is centered at the average of the amplitude values for different cut-offs. For properly renormalized results, we would expect the amplitudes of different regulators to lie within the uncertainty margins. As we can see in  Fig.~\ref{fig:4he_lambda}, this is observed for the LO and one-body NLO corrections. For $\Delta^{(2b)}(\boldq^2)$, on the other hand, the amplitude values span beyond the uncertainty bands. In fact, for the \nlo{4}$^+$ wave function, none of the $\Delta^{(r)}(\boldq^2)$ values fall within the band.

What does this regulator dependence of two-nucleon matrix elements imply? It could indicate a problem with the applied Weinberg power counting for scalar currents. In principle, regulator dependence implies sensitivity to short-distance physics which has to be absorbed by short-range operators, in this case a DM-$N\!N$ interaction. Such terms appear in Weinberg power counting only at N${}^3$LO but might have to be promoted to lower order to ensure order-by-order renormalization. There is by now a lot of evidence that this is the case for neutrinoless double beta decay where a formally subleading contact term must be included at leading order \cite{Cirigliano:2018hja, Cirigliano:2020dmx, Wirth:2021pij, Jokiniemi:2021qqv}. If this is the case here, it implies that including NLO scalar currents for DM-nucleus scattering depends at least on one unknown LEC, not included in existing many-body nuclear structure calculations \cite{Hoferichter:2016nvd,Hoferichter:2018acd}. It would also impact our understanding of the quark mass dependence on nuclear binding energies through the nuclear sigma terms. Interestingly, given sufficiently accurate lattice QCD data on nuclear binding energies as a function of the quark mass, it should be possible to fit the scalar contact counter term to lattice data and determining whether indeed these terms must be promoted to leading order \cite{Beane:2013kca, Chang:2017eiq}. While it is tempting to argue that the missing counter term is not too relevant considering that the two-nucleon matrix elements are found to be small, this is dangerous as there is no guarantee that the finite part of the counter term is of the same size and it could instead be a genuine $\mathcal O(30\%)$ correction.

 On the other hand, it might be that due to some accidental or so far not-understood cancellation, that the NLO two-nucleon matrix elements are small and essentially demoted to N${}^2$LO contributions. The observed regulator dependence is indeed of that size (however, we are only varying the regulators in a small range and the uncertainty might grow with a larger range). If true, it would be interesting to investigate the N${}^2$LO response functions by including higher-order scalar currents derived in Ref.~\cite{Krebs:2020plh}. Unfortunately, these currents come with unknown low-energy constants. 
 
 \begin{figure*}[h!t]
    \centering
    \includegraphics[width=\textwidth]{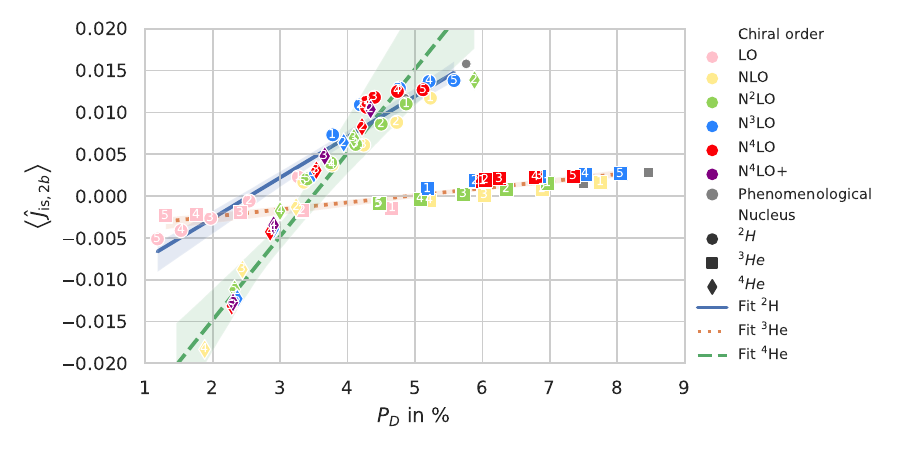}
    \caption{The plots shows the correlation between the isoscalar structure functions and the $D$-wave probability for ${}^2$H, ${}^3$He, and ${}^4$He represented by circle, square, and diamond data points. The data points are assigned different colors corresponding to the chiral order of the wave function. The cut-off $R_i$ used for the calculation is indicated by label $i$ on the data point. For ${}^2$H and ${}^3$He, we have included the cut-off $R_1=0.8 \fm$ and results from phenomenological wave functions  from Ref.~\cite{Korber:2017ery}.}
    \label{fig:d_wave_prob}
\end{figure*}

Finally,  we note that the observed cut-off dependence of our results can be cast in a different light. In Fig.~\ref{fig:d_wave_prob}, we present an observed linear correlation between the two-body matrix element and the $D$-wave probability for ${}^2$H,  ${}^3$He, and ${}^4$He. The $D$-wave probability describes how much of the nuclear wave function involves $l=2$ $N\!N$ states and is a consequence of $S-D$ mixing from the nuclear tensor force. It was pointed out by Friar \cite{Friar:1979zz} already that the $D$-wave probability is not an observable: it can be changed by unitary transformations of the $N\!N$ potential. Therefore, the correlation with the unobservable quantity also points to a potential flaw in the applied power counting.

\section{Conclusions}\label{conclusions}

We have investigated DM scattering of light nuclei using the framework of chiral effective field theory. Our main focus has been on scattering off the ${}^4$He nucleus, a potential target for direct-detection experiments, but we have also investigated ${}^2$H and ${}^3$He nuclei. The nuclear wave functions were calculated from chiral nuclear forces up to fourth order in the chiral expansion. The light nuclei allowed us to calculate bound-state and scattering equations directly with a high degree of accuracy. This allowed us to investigate the uncertainties and convergence of observables for different chiral orders of the wave functions with a high degree of precision. 

We focused on the scalar interaction of DM and quarks. Using \chipt, we calculated the ensuing scalar hadronic current up to NLO (actually N${}^2$LO since no new corrections appear at this order). The NLO current consists of radius and two-body corrections. For ${}^2$H, ${}^3$He,  and ${}^4$He, the NLO currents modify LO results by a few percent only despite larger power-counting expectations. Radius and two-nucleon corrections are found to be of similar size in all three nuclei under investigation. The one-body and radius corrections are in good agreement with earlier work \cite{Korber:2017ery,Andreoli:2018etf}.

Taken at face value, our results imply that NLO corrections are small and the LO response functions can be directly used in the analysis of experiments using light nuclei as targets. Unfortunately, our results for NLO two-nucleon corrections exhibit a clear regulator dependence. While, for ${}^2$H and ${}^3$He we somewhat recover regulator independence when using high-order chiral wave functions, this is not the case for ${}^4$He. Furthermore, in all cases we find a linear correlation between the two-body matrix element and the D-wave probability of the nucleus. Because there are no contact terms at NLO to absorb the regulator dependence, the regulator dependence could indicate that the scalar currents are not properly renormalized. The LO current and NLO radius contributions do not show any such regulator dependence, indicating that all regulator dependence arises from the two-body effects. A potential remedy could be to promote formally subleading DM-nucleon-nucleon interactions to lower order, as is done, for instance, in the description of neutrinoless double beta decay. This would also impact existing predictions for dark matter scattering of heavier nuclei. It could also be that two-nucleon effects should be demoted to N${}^2$LO. More work is needed to clarify this issue for instance by investigating the impact of higher-order scalar currents derived in Ref.~\cite{Krebs:2020plh} in the light nuclei under consideration and beyond.

\begin{acknowledgements}
We thank Martin Hoferichter and Achim Schwenk for useful discussions and encouragement.
JdV acknowledges support from the Dutch Research Council (NWO) in the form of a VIDI grant. This work is further supported in part by the DFG and the NSFC through funds provided to the Sino-German CRC 110 ``Symmetries and the Emergence of Structure in QCD'' (NSFC Grant No. 12070131001, Project-ID 196253076 - TRR 110), and by the MKW NRW under the funding code NW21-024-A.
The numerical calculations were performed on JURECA of the J\"ulich Supercomputing Centre, J\"ulich, Germany.
\end{acknowledgements}


\appendix


\section{Density matrix formalism}
\label{Appx_density}
\subsection{Kinematics and Partial-Wave Decomposition}
We assume that the nucleus has $A$ nucleons, which in the initial state have total angular momentum $J$, spin-projection $M$ onto the $z$-axis, and isospin-projection $M_T$, several isospins $T$ may contribute in general. Here, we will be mostly concerned with $^4$He for which $A=4, J=0, M_T = 0$, the dominant contribution comes from $T=0$, but isospin breaking could in principle introduce small $T=1$  and $2$ contributions to the $^4$He wave function. It turns out that these contributions are not important for the dark matter interactions studied here. 

Additionally, we also assume that the scattering happens in the center of mass frame of the nucleus-DM system. We denote the momentum of the incoming (outgoing) nucleon as $-\boldk$ ($-\boldk'$), and the incoming (outgoing) DM will have the opposite momentum. The momentum of incoming DM  is chosen to lie along the $z$-axis $\boldk = k \vec{e}_z$, which also is the quantization axis for the spin and angular momentum projections.  We furthermore assume that the  scattering is happening in $x$-$z$ plane so that 
$\boldk' = k' \left(  \sin \vartheta \vec{e}_x+ \cos \vartheta  \vec{e}_z \right)$. 
The elastic scattering ensures $\abs{\boldk} = \abs{\boldk'}$, and the transfer momentum into nucleus is $\boldQ = \boldk -\boldk'$.

\begin{figure}[b]
\centering
\begin{tikzpicture}
  \node[circle,draw=black, fill=red, inner sep=0pt,minimum size=8pt,label=left:$1$] (a) at (1,2) {};
  \node[circle,draw=black, fill=red, inner sep=0pt,minimum size=8pt,label=left:$2$] (b) at (0,0) {};
 \node[circle,draw=black, fill=red, inner sep=0pt,minimum size=8pt,label=right:$3$] (c) at (3,-1) {};
   \node[circle,draw=black, fill=red, inner sep=0pt,minimum size=8pt,label=right:$4$] (d) at (4,1) {};

  \node[left] at (0.5,1) {$\vec p_{12}$};
  \node[left] at (3,0.8) {$\vec q_{4}$};
  \node[left] at (2,-0.3) {$\vec p_{3}$};
  
  \draw[line width=2pt,->] (b)--(a);
  \draw[line width=2pt,->] (0.5,1)--(c); 
  \draw[line width=2pt,->] (1.8,0)--(d);
\end{tikzpicture}
\hspace{1cm}
\begin{tikzpicture}
  \node[circle,draw=black, fill=red, inner sep=0pt,minimum size=8pt,label=left:$\displaystyle 1$] (a) at (1,2) {};
  \node[circle,draw=black, fill=red, inner sep=0pt,minimum size=8pt,label=left:$2$] (b) at (0,0) {};
 \node[circle,draw=black, fill=red, inner sep=0pt,minimum size=8pt,label=right:$3$] (c) at (3,-1) {};
   \node[circle,draw=black, fill=red, inner sep=0pt,minimum size=8pt,label=right:$4$] (d) at (4,1) {};
  \node[left] at (0.5,1) {$\vec p_{12}$};
  \node[above] at (2.3,0.6) {$\vec q$};
  \node[right] at (3.5,0.0) {$\vec p_{34}$};
  
  \draw[line width=2pt,->] (b)--(a);
  \draw[line width=2pt,->] (d)--(c);
  \draw[line width=2pt,->] (0.5,1)--(3.5,0); 
\end{tikzpicture}

\caption{\label{fig:jacobi} 3+1 and 2+2 Jacobi coordinates of the four-nucleon system.  }.
\end{figure}
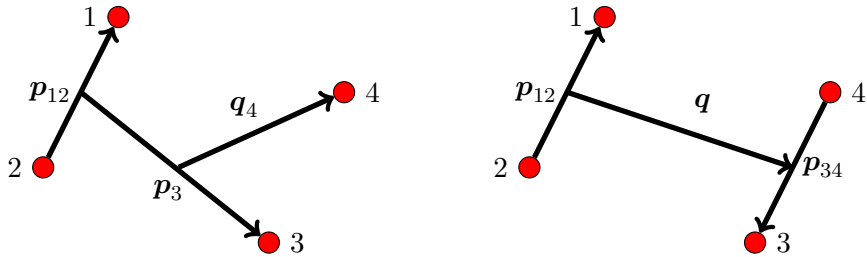

Relative Jacobi coordinates can be defined in various ways by defining different subsystems. 
For the 4N system, except for permutations of 
particles, one finds only two distinct ways for arranging these subclusters, the so-called `3+1' 
and `2+2' coordinates, as depicted in Fig.~\ref{fig:jacobi}. The `3+1' coordinates singles out the last nucleon and provides a relative momentum of this last nucleon with respect to the remaining $A-1$ nucleons. This coordinate will be most useful for the definition of one-nucleon densities below. The `2+2' coordinates  single out the pair of nucleons $1$ and $2$ moving relative to the remaining $A-2$ nucleons. We will later use this coordinate for defining the two-nucleon densities.  

Denoting the nucleon momenta by $\boldk_i$, we can define the Jacobi momentum of the 
spectator particle in the $A-1$+$1$ coordinates as 
\begin{equation}
\boldq_A = \frac{A-1}{A} \boldk_A - \frac{1}{A}\sum_{i=1}^{A-1}\boldk_i = \boldk_A + \frac{1}{A} \boldk
\end{equation}
which, for $^4$He ($A$=4) nucleus, is 
\begin{equation}
\boldq_4 = \frac{3}{4} \boldk_4 - \frac{1}{4}(\boldk_1 + \boldk_2 +\boldk_3) =  \boldk_4 + \frac{1}{4}\boldk.
\label{app_2:jacobi_q4}
\end{equation}
This relation will be important for the implementation of one-body densities. 

For the two-body densities, the relative and total pair momenta of the (12) nucleon system 
\begin{align}
\boldp_{12} &= \frac{1}{2}(\boldk_1 - \boldk_2), & 
\boldk_{12} &= \boldk_1 + \boldk_2
\label{app_2:jacobi_p12}
\end{align}
are the momenta of interest. These can be easiest defined for $A-2$+$2$ coordinates. 
Conveniently, $\boldp_{12}$ is already part of this set of Jacobi coordinates. 
The total pair momentum is related to the relative momentum of the 
(12) subsystem and the remaining nucleons 
\begin{align}
\boldq &= \frac{A-2}{A} \boldk_{12} - \frac{2}{A} \sum_{i=3}^A \boldk_i  = \boldk_{12} + \frac{2}{A} \boldk  = \boldk_{12} + \frac{1}{2} \boldk
\label{app_2:jacobiq}
\end{align}
where the last relation holds for $A=4$. 

We will be focusing on $^4$He for the remaining discussions. We denote the spin-projection state of $^4$He as $\ket{M}$, where we suppress the labels of $JM_T$ and the bound-state energy. This state is the eigenstate for the nucleus at rest, total angular momentum, and its z-component. 
For $^4$He, the total angular momentum of $J=0$ implies that $M=0$. 

The wave function can be expressed in both kinds of coordinates, `3+1' and `2+2'. 
In this work, we use Yakubovsky equations to get the wave functions. In this case, 
both representations of the wave functions are naturally obtained. For calculations that are based on, e.g., the Jacobi no-core shell model \cite{Liebig:2015kwa,Le:2020zdu}, wave functions are usually only 
obtained for `$A$-$1$--$1$' coordinates. However, also in this case, transformation to 
`2+2' coordinates can be easily performed. In following, we will assume that the wave functions 
are given in `3+1' coordinates by 
\begin{equation}
\psi_\al (p_{12}p_3 q_4) = \braket{p_{12}p_3 q_4 \al | M}
\end{equation}
where $\al$ represents the set of angular momentum and isospin quantum numbers defined in 
Eq.~(\ref{eq:psi31}). Similarly, the wave function in `2+2' coordinates reads 
\begin{equation}
\psi_\beta (p_{12}p_{34} q) = \braket{p_{12}p_{34} q \beta | M}. 
\end{equation}
The corresponding set of quantum numbers is defined in Eq.~(\ref{eq:psi22}). 

The labels of outgoing particles are primed. The states $\ket{M}$ must be multiplied by an eigenstate of the nuclear cm momentum operator to conserve the momentum. This will result in the $\delta^{(3)}(\boldk - \boldk'-\boldQ)$. These momentum delta functions are not included in the following calculations  to shorten the notation. 

\subsection{One-Body Density}

Assuming that the active nucleon is the fourth one, the relevant matrix element for one-body operator reads 
\begin{align}
    &\bra{\boldk_4'} {\bra{\lfourpc \mlfourpc}}\bra{\sfourpc \msfourpc} \bra{\tfourpc \mtfourpc }\hat{O}_4(\boldk, \boldQ) \nonumber \\ 
    & \hspace{3.5cm} \ket{\tfourc \mtfourc} \ket{\sfourc \msfourc} {\ket{\lfourc \mlfourc}} \ket{\boldk_4}
    \nn \\
    &=
    \delta_{m^{t'}_4,m^t_4}\delta^{(3)}(\boldk_4' -\boldk_4-\boldQ) O_4(m^{s'}_4m^s_4m^t_4;\boldk_4; \boldk,\boldQ).
    \label{app_2:one_body_O4hat}
\end{align}
Here $m^t_4$ is conserved because the DM interaction does not change the charge of the struck nucleon. 
For the one-body density, we allow for a dependence of the operator on the momentum $\boldk_4$ 
of the active nucleon which is necessary to take boost corrections into account as they appear in subleading 
operators. This explicit dependence on the momentum of the active nucleon is included in
terms of a multipole expansion which we  truncate at  order $K_{\rm max}$. 
\begin{eqnarray}
 & & O_4( \msfourpc \msfourc \mtfourc;\boldk_4; \boldk,\boldQ) \equiv \sum_{K=0}^{K_{\rm max}}\sum_{\kappa=-K}^{K}
\sqrt{\frac{4\pi}{2K+1}}
(k_4)^K  \nonumber \\
& &  \hspace{2cm}\times Y_{K\kappa}(\hat k_4) \Tilde{O}_4(\msfourpc \msfourc \mtfourc;K\kappa;\boldk,\boldQ),
\label{app_2:one_body_multipole}
\end{eqnarray}
where $\hat{k}_4$ is the unit vector of $\boldk_4$. In practice, we only use $K_{\rm max}=1$.

Now let us look at the matrix element of $\hat O_4$ that we want to calculate
\begin{align}
   & \bra{M'} \hat O_4(\boldk, \boldQ) \ket M = 
    \sum_{\al \al'} \nn \\ & \int \dd p_{12}\ p_{12}^2\ \dd p_3\ p_3^2\ \dd q_4\ q_4^2\ \dd p_{12}'\ p_{12}^{'2}\ \dd p_3'\ p_3^{'2}\ \dd q_4'\ q_4^{'2}\ \nn \\ & \times \psi^\dagger_{\al'}(p_{12}'p_3' q_4')\psi_{\al}(p_{12}p_3 q_4)
    \nn \\
    & \times 
    \left\langle p_{12}' p_{3}' q_{4}', \left[ [\lpab \spab] \jpab (\lthreepc \sthreepc) \Ithreepc] j_{3}' (\lfourpc \sfourpc) \Ifourpc \right]  J'M', \right. \nn \\ & \hspace{2cm} \left. \left[ (\tpab \tthreepc) \tau_3' \tfourpc \right]  T'M_T' \right| \hat O_4(\boldk, \boldQ)
    \nn \\
    & \hspace{1cm}
    \left| p_{12} p_{3}, \left[ [\lab \sab] \jab (\lthreec \sthreec) \Ithreec] j_{3} (\lfourc \sfourc) \Ifourc \right] JM, \right.  \nn \\ & \hspace{2cm}\left. \left[ (\tab \tthreec) \tau_3 \tfourc \right]  TM_T \right\rangle .
    \label{app_2:MO3M_1N_one}
\end{align}
 Using Clebsch-Gordan coefficients $\CG{j_1}{j_2}{j}{m_1}{m_2}{m}$  we can explicitly decompose $\al$ so as to separate the spin-isospin
quantum numbers of the active nucleon 4  from those of the other three nucleons:
\begin{align}
   & \bra{M'}\hat O_4(\boldk, \boldQ) \ket M
    =  \sum_{\al \al'} \nn \\
   &  \int \dd p_{12}\ p_{12}^2\ \dd p_3\ p_3^2\ \dd q_4\ q_4^2\ \dd p_{12}'\ p_{12}^{'2}\ \dd p_3'\ p_3^{'2} \ \dd q_4'\ q_4^{'2}\nn \\ & \times \psi^\dagger_{\al'}(p_{12}'p_3'q_4')\psi_{\al}(p_{12}p_3 q_4)
    \nn \\
   & \times 
   \sum_{\mfourpc} \CG{j_3'}{\Ifourpc}{J'}{(M'-\mfourpc)}{\mfourpc}{M'} \nn \\ & \times 
   \sum_{\mtfourpc} \CG{\tau_3'}{\tfourpc}{T'}{(M_T'-\mtfourpc)}{\mtfourpc}{M_T'}
    \nn \\
   & \times 
   \sum_{\mfourc} \CG{j_3}{\Ifourc}{J}{(M-\mfourc)}{\mfourc}{M} \nn \\ & \times 
   \sum_{\mtfourc} \CG{\tau_3}{\tfourc}{T}{(M_T-\mtfourc)}{\mtfourc}{M_T}
    \nn \\
   & \times 
   \left\langle p_{12}'p_{3}',  [\lpab \spab] \jpab (\lthreepc \sthreepc) \Ithreepc] j_{3}'(M'-\mfourpc), \right. \nn \\ & \left. \hspace{2cm}(\tpab \tthreepc) \tau_3' (M_T'-\mtfourpc) \right|  \nn \\ 
   & \qquad \left| p_{12} p_{3}, [\lab \sab] \jab (\lthreec \sthreec) \Ithreec] j_{3}(M-\mfourc), \right. \nn \\ & \hspace{2cm} \left. (\tab \tthreec) \tau_3 (M_T-\mtfourc) \right\rangle 
    \nn \\
   & \times 
   \braket{q_4', (\lfourpc \sfourpc)\Ifourpc \mfourpc, \tfourpc \mtfourpc|\hat O_4(\boldk,\boldQ)|q_4, (\lfourc \sfourc)\Ifourc \mfourc, \tfourc \mtfourc }
    \label{app_2:MO3M_1N_two}
\end{align}
Here we used identities from the spin projections: $M= m_4+m_{123}, M_T= m^t_4+m^t_{123}, \ldots$. Now we expand the last term in the above equation using Eq.~\eqref{app_2:one_body_O4hat}, the momentum conserving delta function can be rewritten using Eq.~\eqref{app_2:jacobi_q4} as
\begin{equation}
\delta^{(3)}(\boldk_4'-\boldk_4-\boldQ) =\delta^{(3)}(\boldq_4'-\boldq_4 -\frac{3}{4}\boldQ) \, .
\end{equation}
We further expand the last line of Eq.~\eqref{app_2:MO3M_1N_two} by inserting the solid angles $\hat q^{}_4, \hat q_4'$ of the third particles' momentum. 
The state of the active nucleon is further decoupled into an orbital and 
spin part which introduces summations of spin projections $\msfourpc$ and  $\msfourc$ . Then we use the result in Eq.~\eqref{app_2:one_body_multipole}, and perform the integrations and summations corresponding to the appropriate delta functions to arrive at the final form
\begin{align}
    & \bra{M'}\hat O_4(\boldk, \boldQ) \ket M
    =
    \sum_{K=0}^{K_{\rm max}}\sum_{\kappa=-K}^{K} \nn \\ & \sum_{\substack{m_4^{s'}m_4^s\\ m^t_4}}
    \Tilde{O}_4(\msfourpc \msfourc \mtfourc;K\kappa;\boldk,\boldQ) \rho^{K\kappa; \mtfourc M_T,M'M}_{\msfourpc \msfourc}(\boldk, \boldQ) \, ,
\end{align}
where we define the one-body (transition) density by summing over those quantum numbers that are not involved in the interaction:
\begin{align}
& \rho^{K\kappa; \mtfourc M_T,M'M}_{\msfourpc \msfourc}(\boldk, \boldQ)
    \nn \\
    &= \sum_{\al \al'} \int \dd p_{12}\ p_{12}^2\ \dd p_3\ p_3^2\ \dd q_4\ q_4^2\  
    \nn \\
  & \times \delta_{\lpab \lab}\, \delta_{\spab \sab}\, \delta_{\jpab \jab}\,  \delta_{\tpab \tab}\, \delta_{\lthreepc \lthreec }\,
  \delta_{\Ithreepc \Ithreec }\,  \delta_{j_3' j_3}
  \delta_{\tau_3'  \tau_3 } \,   \delta_{M_T'M_T}\,   \nn \\
   & \times 
   \sum_{\mfourc} \CG{j_3'}{\Ifourpc}{J'}{(M-\mfourc)}{M'-M+\mfourc}{M'} \nn \\ 
   & \times \CG{\tau_3'}{\tfourpc}{T'}{(M_T-\mtfourc)}{\mtfourc}{M_T}
    \nn \\
   & \times 
    \CG{j_3}{\Ifourc}{J}{(M-\mfourc)}{\mfourc}{M}
    \nn \\ & \times \CG{\tau_3}{\tfourc}{T}{(M_T-\mtfourc)}{\mtfourc}{M_T}
    \nn \\
& \times 
    \CG{\lfourpc}{\sfourpc}{\Ifourpc}{(M'-M+\mfourc - \msfourpc)}{\msfourpc}{(M'-M+\mfourc)} \nn \\ & \times 
    \CG{\lfourc}{\sfourc}{\Ifourc}{(\mfourc-\msfourc)}{\msfourc}{\mfourc}
    \nn \\
& \times
    \int \dd \hat q_4
    Y^{\dagger}_{\lfourpc (M'-M+\mfourc - \msfourpc)}(\widehat{\boldq_4 + \tfrac{3}{4} \boldQ}) 
    Y_{\lfourc (\mfourc - \msfourc)}(\hat \boldq_4)
    \nn \\
& \times
    \sum_{K=0}^{K_{\rm max}}\sum_{\kappa=-K}^{K}
    \sqrt{\frac{4\pi}{2K+1}}
    |\boldq_4-\tfrac{\boldk}{4}|^K Y_{K\kappa}(\widehat{\boldq_4-\tfrac{\boldk}{4}})  
    \nn \\ & \times \psi^{\dagger}_{\al'}(p^{}_{12}p^{}_{3}\abs{\boldq_4 + \tfrac{3}{4} \boldQ}) \psi_{\al}(p_{12}p_{3}q_{4}) 
\end{align}
where $\widehat{\vec a + \vec b}$ denotes the direction of $\vec a + \vec b$. 
This definition separates the calculation of the operator from the 
calculation of the wave functions. Based on this definition, we can define 
transition densities for arbitrary nuclear wave functions that can then 
be combined with separately calculated matrix elements of operators. 

\subsection{Two-Body Density}
In this work, we take into account one- and two-nucleon 
operators contributing to elastic DM scattering. For the application 
of two-nucleon operators, the relevant matrix element is given by
\begin{align}
    &\bra{\boldp_{12}' \boldk_{12}'} \bra{\spab \mspab} \bra{\tpab \mtpab}  \nn \\ & \hspace{2.5cm} \hat O_{12}(\boldk, \boldQ) \ket{\tab \mtab} \ket{\sab \msab} \ket{{\boldp_{12} \boldk_{12}}}
    \nn \\ 
    & =
    \delta^{(3)}(\boldk_{12}'-\boldk_{12}-\boldQ) \delta_{\mtpab \mtab} \nn \\
    & \qquad O_{12} (\spab \tpab \mspab \sab \tab \mtab;\boldp_{12}',\boldp_{12};\boldQ).
\end{align}
For these operators, that generally contribute only at higher orders, it will not be necessary to take boost corrections into account. 
Therefore, we assume that the operator does not have an explicit dependence 
on $\vec k_{12}$. 
The two-nucleon operator is usually represented in terms of partial-wave states, which can be written as,
\begin{align}
    & \braket{\al_{12}' p_{12}'\boldk_{12}'|\hat O_{12}|\al_{12} p_{12}\boldk_{12}}
    \nn \\ & = \delta_{\mtpab \mtab} \delta^{(3)}(\boldk_{12}'-\boldk_{12}-\boldQ)
    \nn \\
    & \times
    \sum_{\mspab \msab}
    \CG{\lab}{\sab}{\jab}{(\mab-\msab)}{\msab}{\mab} \nn \\ & \qquad \times 
    \CG{\lpab}{\spab}{\jpab}{(\mpab-\mspab)}{\mspab}{\mpab}
    \nn \\
    & \times
    \int \dd \hat p_{12}' \dd \hat p_{12} Y^{\dagger}_{\lpab(\mpab-\mspab)}(\hat p_{12}') Y_{\lab (\mab-\msab)}(\hat p_{12})
    \nn \\
    & \times
    O_{12} (\spab \tpab \mspab \sab \tab \msab \mtab; \boldp_{12}',\boldp_{12}; \boldQ)
    \nn \\
    & \equiv
    \delta_{\mtpab \mtab} \delta^{(3)}(\boldk_{12}'-\boldk_{12}-\boldQ) O_{12}^{\al_{12}'\al_{12}}(p_{12}',p_{12};  \boldQ ).
\end{align}
For the calculation of matrix elements of such an operator, it is advisable 
to use coordinates that separate the two-nucleon cluster from a cluster 
of the remaining nucleons. For an $A=4$ system, such coordinates 
are usually labeled as `2+2' coordinates and are defined in Eq.~(\ref{eq:psi22}). The scattering process changes the momenta within 
the subcluster (12) and the relative momentum $\boldq$ of this subcluster 
and the remaining nucleons. Using Eq.~\eqref{app_2:jacobiq} and Eq.~\eqref{app_2:jacobi_p12} along with $\sum_i \boldk_i = -\boldk$, we can rewrite the delta function as, 
\begin{equation}
 \delta^{(3)}(\boldk_{12}'-\boldk^{}_{12}-\boldQ) =  \delta^{(3)}(\boldq -\boldq'+\frac{1}{2}\boldQ) \, . 
 \end{equation}
 We can use this delta function to integrate $\boldq$ when calculating the matrix element. Now we can write the two-body matrix element following the same procedure as one-body matrix element and we get the factorization, 
 \begin{eqnarray}\label{appA:two_body_density}
 & & \bra{M'}\hat O_{12}(\boldq) \ket M \nn \\
 &  &  \equiv  
 \sum_{\al_{12}'\al^{}_{12}} 
 \int \dd p_{12} p_{12}^2 \dd p_{12}' \dd p_{12}^{'2} \ O_{12}^{\al_{12}'\al_{12}}(p_{12}',p_{12};\boldQ) \nn \\ & & \qquad \qquad \rho_{\al_{12}'\al_{12}}^{M_T, M',M}(p_{12}',p^{}_{12};\boldQ)
 \end{eqnarray}
 where we define the two-body (transition) density as:
\begin{align}
   & \rho_{\al_{12}'\al^{}_{12}}^{M_T, M',M} (p_{12}' p^{}_{12};\boldQ) 
   \nn \\
   & =
   \sum_{\al'(\al_{12}')\al(\al^{}_{12})} \sum_{m_I}
    \CG{I}{j_{34}}{J}{m_I}{M-m_I}{M} \nn \\ & \qquad \times 
   \CG{I'}{j_{34}}{J'}{m_I+M'-M}{M-m_I}{M'}
    \delta_{j_{34}^{} j_{34}^{'}}\nn \\
& \times \CG{\jab}{\lambda}{I}{\mab}{(m_I-\mab)}{m_I} \nn \\ & \times 
   \CG{\jpab}{\lambda'}{I'}{\mpab}{(M'-M+m_I-\mpab)}{(m_I+M'-M)} \nn \\
      & \times
   \CG{\tab}{t_{34}}{T}{\mtab}{(M_T-\mtab)}{M_T}
   \nn \\ & \times \CG{\tpab}{t_{34}'}{T'}{\mtab}{(M_T-\mtab)}{M_T}  
   \nn \\
   & \times
   \int \dd q' q^{'2} \int \dd \hat q' \int \dd p_{34}' p_{34}^{'2} \  \psi^{\dagger}_{\al'}(p_{12}'p_{34}'q') \nn \\ & \hspace{4cm} \times \psi_{\al}(p^{}_{12}p_{34}'\abs{\boldq'+\tfrac{1}{2}\boldQ}) \nn \\ 
   & \times     Y^{\dagger}_{\lambda' (M'-M+m_I'-m_{12}')}(\hat q') Y_{\lambda(m_I-m_{12})}(\widehat{\boldq'+\tfrac{1}{2}\boldQ}).
\end{align}

\section{Partial wave decomposition}
\label{pwd}
In this section, we give the partial wave decomposition of two-body operator $O_{12}^{\alpha_{12}'\alpha^{}_{12}}(p_{12}',p^{}_{12};\boldQ)$ in \eqref{appA:two_body_density}. For notational convenience, we will drop the subscript (12) from now on. We can decompose the operator into total-spin part and isospin part
\begin{align}
    & O^{\alpha'\alpha}_{\mc'\mc}(p',p;\boldQ) \nn \\ 
    & =
    \left(
        \sum_{\xi =0}^{\infty} \sum_{ |m_{\xi}| \leq \xi}
        \CG{j}{\xi}{j'}{m_j}{m_{\xi}}{m_j'}
        \mathcal{O}_{ (\alpha' \alpha)\xi m_{\xi},\mc'\mc }^{(\text{spin})}
    \right)
    \nonumber 
    \\ & \hspace{1cm}
  \times 
	\left(
        \sum_{\tau =0}^{2} \sum_{ |m_{\tau}| \leq \tau}
        \CG{t}{\tau}{t'}{m_t}{m_{\tau}}{m_t}
        \mathcal{O}_{ (t' t)\tau m_{\tau} }^{(\text{iso})}
    \right)\,,
\end{align}
with 
\begin{align}
    & \mathcal{O}_{ (\alpha' \alpha)\xi m_{\xi},\mc'\mc }^{(\text{spin})}
    =
    \sqrt{\hat j \hat l' \hat s' \hat \xi}
    \sum_{\lambda \sigma}
    \left(
    	\begin{array}{ccc}
    		j & \xi & j' \\
    		l & \lambda & l' \\
    		s & \sigma & s'
    	\end{array}
    \right)
    \nn \\ & \hspace{3cm} \mathcal{O}_{ ((l'l)\lambda,  (s's)\sigma)\xi m_\xi,\mc'\mc } (p',p,\boldQ)\,,
\end{align}
and
\begin{align}
    \mathcal{O}_{ (t' t)\tau m_{\tau} }^{(\text{iso})}
    =
    \frac{\hat \tau}{ \hat t' }
    \sum_{m_t}
    \CG{t}{\tau}{t'}{m_t}{m_{\tau}}{m_t}
    \langle t' m_t | O^{(\text{iso})} | t m_t  \rangle\,,
    \label{eq:pwd_isospin}
\end{align}
where $\mc (\mc')$ is the spin quantum number of the incoming (outgoing) dark matter, $\hat x \equiv 2 x +1$, and \\ $\CG{j_1}{j_2}{j}{m_1}{m_2}{m}$ are the Clebsch-Gordan coefficients. The partial wave decomposition of the isospin part is given in Table \ref{tab:Ospin_list}.

\begin{table}[t!]
    \centering
    \begin{tabular}{|cccc|c|}
     \hline 
     $t'$ &  $t$ & $\tau$ & $m_\tau $ & $\tilde{\mathcal{O}}_{(t't)\tau m_\tau}^{(\text{iso})}$\\
     \hline
     0 & 0 & 0 & 0& -3   
     \\
     1 & 1 & 0 & 0 & 1 \\
    \hline
\end{tabular}
    \caption{Partial wave decomposition of isospin operator $O^{(\text{iso})}$ as given in Eq. \eqref{eq:pwd_isospin} for non-zero channels.}
    \label{tab:Ospin_list}
\end{table}

The decomposed operator can be factorized in an angular momentum- and spin-dependent parts
\begin{align}
& \mathcal{O}_{ ((l'l)\lambda,  (s's)\sigma)\xi m_\xi,\mc'\mc }(p',p;\boldQ)
\nonumber\\
&\equiv
    \sum_{m_\lambda m_\sigma}
    \CG{\lambda}{\sigma}{\xi }{m_\lambda}{m_\sigma }{m_\xi}
\nonumber    
\\
& \hspace{1cm} \times
     \int \dd \theta'\int \dd \phi'
    \int \dd \theta\int \dd \phi 
    \mathcal{O}_{(l'l)\lambda m_\lambda}^{(L)} ( \theta', \phi', \theta,\phi) \nn \\ & \hspace{4cm} \times 
    \mathcal{O}_{(s's)\sigma m_\sigma,\mc'\mc}  ^{(S)} (\boldp', \boldp; \boldQ)\,,
\end{align}
the angular momentum-dependent part is a function of spherical harmonics while the spin-dependent part evaluates the remaining matrix-element
\begin{align}
    & \mathcal{O}_{(l'l)\lambda m_\lambda}^{(L)}(\theta', \phi', \theta,\phi)
    \equiv
    \frac{\hat \lambda }{\hat l'}
    \sum_{m_l' m_l}
    \CG{l}{\lambda}{l'}{m_l}{m_\lambda}{m_l'}
    \nn \\ & \hspace{2cm} \times 
    Y_{l' m_l'}^*(\theta', \phi')
    Y_{l  m_l }  (\theta,\phi)
    \label{eq:OL_before}
\end{align}
and
\begin{align}
    & \mathcal{O}_{(s's)\sigma m_\sigma,\mc'\mc}^{(S)}(\boldp', \boldp; \boldQ) \nn \\ \qquad 
    & \equiv
    \frac{\hat \sigma }{\hat s'}
    \sum_{m_s' m_s}
    \CG{s}{\sigma }{s'}{m_s}{m_\sigma}{m_s'}
    \nn \\ & \hspace{1.7cm}
    \langle s' m_s' \mc'| O(\boldp', \boldp; \boldQ) | s m_s\mc \rangle \, . 
    \label{eq:Os_spin}
\end{align}
Conservation laws which simplify the angular-integration can be applied depending on the spin matrix-elements of the operator $\mathcal{O}_{(s's)\sigma m_\sigma,\mc'\mc}^{(S)}$.

\subsection{Isoscalar two-body correction}

For two body operators, we can write the Jacobi momenta in terms of nucleon momentum $\boldk_i, \boldk_i'$ as
\begin{align}
    \boldp_{12} &= \frac{\boldk_1-\boldk_2}{2}\,, &
    \boldp_{12}' &= \frac{\boldk_1'-\boldk_2'}{2}\,,
    \\
\boldq_1 &= \frac{\boldQ}{2}+ \boldp_{12}-\boldp_{12}'\,, &
    \boldq_2 &= \frac{\boldQ}{2}- \boldp_{12}+\boldp_{12}'\,.
\end{align}

For scalar dark matter interactions, the spin matrix element in \eqref{eq:Os_spin} is proportional to \\ $\abs{\boldk}\equiv \abs{\boldp^{}_{12} + \tilde\alpha \boldp_{12}' + \tilde\beta \boldQ}$ and $m_\xi=0$. In the structure $\abs{\boldk}$, azimuthal angles appear as $\phi_{p^{}_{12}}-\phi_{p_{12}'}$ (using $\boldQ = Q \vec{e}_z$). It is convenient to use the following transformation
\begin{align}
    \int \dd \phi_{p^{}_{12}}\int \dd \phi_{p_{12}'} \rightarrow \int \dd \varphi \int \dd \Phi\,,
\end{align}
with $\varphi =  \phi_{p^{}_{12}} - \phi_{p_{12}'}, \Phi = \tfrac{\phi^{}_{p_{12}} + \phi_{p_{12}'}}{2}$. This transformation modifies the spherical harmonics in Eq. \eqref{eq:OL_before} as
\begin{align}
    & Y_{l' m_l'}^*(\theta', \phi')
    Y_{l  m_l }  (\theta,\phi) \nn \\ & \qquad 
    \rightarrow
    Y_{l' m_l'}^*(\theta', 0)
    Y_{l  m_l }  (\theta,0)
    e^{i m_l \varphi}e^{i m_\lambda \left(\frac{\varphi}{2}-\Phi\right)} \, ,
\end{align}
where we used $m_l + m_\lambda = m_l' $ and $m_\xi =0$, thereby $m_\lambda = - m_\sigma$. This allows us to absorb one integration in the spin matrix-element
\begin{align}
&\mathcal{O}_{ ((l'l)\lambda,  (s's)\sigma)\xi m_\xi,\mc'\mc } (p_{12}',p_{12},Q)
\nonumber\\
& \hspace{0.5cm}\equiv
    \sum_{m_\lambda m_\sigma}
    \CG{\lambda}{\sigma}{\xi }{m_\lambda}{m_\sigma }{m_\xi}
\nonumber    
\\
& \hspace{1.5cm} \times
     \int \dd \theta'
    \int \dd \theta\int \dd \varphi \ 
    \tilde{\mathcal{O}}_{(l'l)\lambda m_\lambda}^{(L)} ( \theta_{p_{12}'}, \theta_{p_{12}},\varphi) \nn \\ & \hspace{3cm} \times 
    \tilde{\mathcal{O}}_{(s's)\sigma m_\sigma,\mc'\mc}  ^{(S)} (p_{12}', p_{12}, Q,\varphi)\,,
\end{align}
with
\begin{align}
    & \tilde{\mathcal{O}}_{(l'l)\lambda m_\lambda}^{(L)}(\theta_{p_{12}'}, \theta_{p_{12}},\varphi) \nn \\ & \quad 
    \equiv
    \frac{\hat \lambda }{\hat l'}
    \sum_{m_l' m_l}
    \CG{l}{\lambda}{l'}{m_l}{m_\lambda}{m_l'}
    \nn \\ & \hspace{1cm} \times 
    Y_{l' m_l'}^*(\theta_{p_{12}'}, 0)
    Y_{l  m_l }  (\theta_{p_{12}},0)e^{im_l\varphi } \,,
    \\
    & \tilde{\mathcal{O}}_{(s's)\sigma m_\sigma,\mc'\mc}^{(S)}(p_{12}',p_{12},Q,\varphi)
     \nonumber \\
    & \equiv
    \frac{\hat \sigma }{\hat s'}
    \sum_{m_s' m_s}
    \CG{s}{\sigma }{s'}{m_s}{m_\sigma}{m_s'}
    \nn \\ & \hspace{0.7cm}
    \int \dd\Phi
    \langle s' m_s' \mc'| O(\boldp_{12}', \boldp_{12}, \boldQ) | s m_s\mc \rangle e^{im_\sigma(\Phi - \tfrac{\varphi}{2})}\,.
\end{align}

For scalar dark matter scattering, the last expression is either $\Phi$ independent or easy enough to perform the integration analytically. This significantly reduces the numerical scaling and is eventually used for computation. 

\begin{table}[t!]
    \centering
    \begin{tabular}{|cccc|c|}
     \hline 
     $s'$ &  $s$ & $\sigma$ & $m_\sigma$ & $\tilde{\mathcal{O}}_{(s's)\sigma m_\sigma,\mc'\mc}^{(S)}(p',p'q,\varphi)$\\
     \hline 
    0  & 0 & 0 & 0 & -2$\pi (\alpha - 2 \gamma)$\\
    0  & 1 & 1 & -1 & $\sqrt 6 \pi \beta Q$\\
    0  & 1 & 1 & 1 & $\sqrt 6 \pi \beta^* Q$\\
    1  & 0 & 1 & -1 & $\sqrt 2 \pi \beta Q$\\
    1  & 0 & 1 & 1 & $\sqrt 2 \pi \beta^* Q$\\
    1  & 1 & 0 & 0 & $\frac{2}{3}\pi(\alpha -2\gamma) $\\
    1  & 1 & 2 & -2 & $2 \sqrt{\frac{5}{3}}\pi \beta^2$\\
    1  & 1 & 2 & -1 & $4\sqrt{\frac{5}{3}}\pi \beta \delta $\\
    1  & 1 & 2 & 0 & $\frac{1}{3} \sqrt{10} \pi (\alpha + 3 \delta^2 -3 \abs{\beta}^2-2\gamma -\frac{3}{4}Q^2)$\\
    1  & 1 & 2 & 1 & $-4 \sqrt{\frac{5}{3}}\pi \beta^* \delta$\\
    1  & 1 & 2 & 2 & $2\sqrt{\frac{5}{3}}\pi (\beta^*)^2$\\
    \hline
\end{tabular}
    \caption{Partial wave decomposition of spin operator in Eq. \eqref{eq:dm_spin_op} for non-zero channels. The coefficients $\alpha, \beta, \gamma,\delta$ are defined in Eq. \eqref{eq:tab_coeff_def}.}
    \label{tab:Os_list}
\end{table}

Here we perform the above decomposition for Diagram \ref{fig:isoscalar_diag}(c) given by Eq. \eqref{NLO2body}. For simplicity, we are omitting the overall $\mpi^2 \left(\frac{g_A}{2f_\pi}\right)^2$ factor and the $\Phi$ independent denominator $(\boldq_1^2 + \mpi^2) (\boldq_2^2 + \mpi^2)$. The spin-momentum decomposition for this operator  
\begin{align}
& \tilde{\mathcal{O}}_{(s's)\sigma m_\sigma,\mc'\mc}^{(S)}(p_{12}',p_{12},Q,\varphi) \nonumber \\
    &\quad  =
    \frac{\hat \sigma }{\hat s'}
    \sum_{m_s' m_s}
    \CG{s}{\sigma }{s'}{m_s}{m_\sigma}{m_s'}
    \nn \\ & \hspace{0.6cm} \times 
    \int \dd \Phi
    \langle s' m_s' |- (\boldsigma_1 \cdot \boldq_1)(\boldsigma_2 \cdot \boldq_2) | s m_s \rangle e^{im_\sigma(\Phi - \tfrac{\varphi}{2})}\,
    \label{eq:dm_spin_op}
\end{align}
is given in Table \ref{tab:Os_list}. The coefficient used in the table are
\begin{align}
    \alpha &= p_{12}^2 + p_{12}'^2-\left(\frac{Q}{2}\right)^2 \nn \\ 
    \beta &= e^{i\varphi} p_{12} \sin (\theta_{p_{12}})-p_{12}'\sin \theta_{p_{12}'}\,,
    \nonumber
    \\
    \gamma &= p_{12} p_{12}' (\cos \theta_{p_{12}} \cos \theta_{p_{12}'} + \cos \varphi \sin \theta_{p_{12}} \sin \theta_{p_{12}'})\,, \nn \\ 
    \delta &= p_{12} \cos \theta_{p_{12}} - p_{12}' \cos \theta_{p_{12}'}\,.
\label{eq:tab_coeff_def}
\end{align}

The momenta $\boldq_{1,2}$ and the denominator are given by
\begin{align}
    & \boldq_{1,2}^2 = \alpha -2\gamma \pm \delta Q + \frac{1}{2}Q^2 \nn \\ 
    & \quad \Rightarrow (\boldq_1^2 + \mpi^2) (\boldq_2^2 + \mpi^2) \nn \\ & \hspace{1.5cm} = \left(\alpha -2\gamma + \frac{1}{2} Q^2 + \mpi^2\right)^2 -\delta^2 Q^2\,.
\end{align}
This expression is spin independent. All these structures are combined and convoluted with the density matrix which is used to  numerically calculate the scattering matrix elements.

\bibliographystyle{h-physrev3}
\bibliography{mybib}

\begin{thebibliography}{10}

\bibitem{Bertone:2016nfn}
G.~Bertone and D.~Hooper,
\newblock Rev. Mod. Phys. {\bf 90}, 045002 (2018), 1605.04909.

\bibitem{Battaglieri:2017aum}
M.~Battaglieri {\em et~al.},
\newblock {US Cosmic Visions: New Ideas in Dark Matter 2017: Community Report},
\newblock in {\em {U.S. Cosmic Visions: New Ideas in Dark Matter}}, 2017,
  1707.04591.

\bibitem{Hertel:2018aal}
S.~A. Hertel, A.~Biekert, J.~Lin, V.~Velan, and D.~N. McKinsey,
\newblock Phys. Rev. D {\bf 100}, 092007 (2019), 1810.06283.

\bibitem{Schumann:2019eaa}
M.~Schumann,
\newblock J. Phys. G {\bf 46}, 103003 (2019), 1903.03026.

\bibitem{Fan:2010gt}
J.~Fan, M.~Reece, and L.-T. Wang,
\newblock JCAP {\bf 11}, 042 (2010), 1008.1591.

\bibitem{Fitzpatrick:2012ix}
A.~L. Fitzpatrick, W.~Haxton, E.~Katz, N.~Lubbers, and Y.~Xu,
\newblock JCAP {\bf 02}, 004 (2013), 1203.3542.

\bibitem{Cirigliano:2012pq}
V.~Cirigliano, M.~L. Graesser, and G.~Ovanesyan,
\newblock JHEP {\bf 10}, 025 (2012), 1205.2695.

\bibitem{Klos:2013rwa}
P.~Klos, J.~Men\'endez, D.~Gazit, and A.~Schwenk,
\newblock Phys. Rev. D {\bf 88}, 083516 (2013), 1304.7684,
\newblock [Erratum: Phys.Rev.D 89, 029901 (2014)].

\bibitem{Hoferichter:2015ipa}
M.~Hoferichter, P.~Klos, and A.~Schwenk,
\newblock Phys. Lett. B {\bf 746}, 410 (2015), 1503.04811.

\bibitem{Brod:2017bsw}
J.~Brod, A.~Gootjes-Dreesbach, M.~Tammaro, and J.~Zupan,
\newblock JHEP {\bf 10}, 065 (2018), 1710.10218.

\bibitem{Richardson:2021liq}
T.~R. Richardson, X.~Lin, and S.~T. Nguyen,
\newblock Phys. Rev. C {\bf 106}, 044003 (2022), 2110.15959.

\bibitem{DEramo:2016gos}
F.~D'Eramo, B.~J. Kavanagh, and P.~Panci,
\newblock JHEP {\bf 08}, 111 (2016), 1605.04917.

\bibitem{Bishara:2018vix}
F.~Bishara, J.~Brod, B.~Grinstein, and J.~Zupan,
\newblock JHEP {\bf 03}, 089 (2020), 1809.03506.

\bibitem{Prezeau:2003sv}
G.~Prezeau, A.~Kurylov, M.~Kamionkowski, and P.~Vogel,
\newblock Phys. Rev. Lett. {\bf 91}, 231301 (2003), astro-ph/0309115.

\bibitem{Korber:2017ery}
C.~K\"orber, A.~Nogga, and J.~de~Vries,
\newblock Phys. Rev. C {\bf 96}, 035805 (2017), 1704.01150.

\bibitem{Hoferichter:2016nvd}
M.~Hoferichter, P.~Klos, J.~Men\'endez, and A.~Schwenk,
\newblock Phys. Rev. D {\bf 94}, 063505 (2016), 1605.08043.

\bibitem{Beane:2013kca}
S.~R. Beane, S.~D. Cohen, W.~Detmold, H.~W. Lin, and M.~J. Savage,
\newblock Phys. Rev. D {\bf 89}, 074505 (2014), 1306.6939.

\bibitem{Andreoli:2018etf}
L.~Andreoli, V.~Cirigliano, S.~Gandolfi, and F.~Pederiva,
\newblock Phys. Rev. C {\bf 99}, 025501 (2019), 1811.01843.

\bibitem{Cirigliano:2018hja}
V.~Cirigliano {\em et~al.},
\newblock Phys. Rev. Lett. {\bf 120}, 202001 (2018), 1802.10097.

\bibitem{SPICEHeRALD:2021jba}
SPICE/HeRALD, A.~Biekert {\em et~al.},
\newblock Phys. Rev. D {\bf 105}, 092005 (2022), 2108.02176.

\bibitem{Liao:2021npo}
J.~Liao {\em et~al.},
\newblock (2021), 2103.02161.

\bibitem{Arcadi:2019lka}
G.~Arcadi, A.~Djouadi, and M.~Raidal,
\newblock Phys. Rept. {\bf 842}, 1 (2020), 1903.03616.

\bibitem{Hu:2021awl}
B.~S. Hu {\em et~al.},
\newblock Phys. Rev. Lett. {\bf 128}, 072502 (2022), 2109.00193.

\bibitem{ParticleDataGroup:2022pth}
Particle Data Group, R.~L. Workman {\em et~al.},
\newblock PTEP {\bf 2022}, 083C01 (2022).

\bibitem{Brantley:2016our}
D.~A. Brantley {\em et~al.},
\newblock (2016), 1612.07733.

\bibitem{Hoferichter:2015dsa}
M.~Hoferichter, J.~Ruiz~de Elvira, B.~Kubis, and U.-G. Mei\ss{}ner,
\newblock Phys. Rev. Lett. {\bf 115}, 092301 (2015), 1506.04142.

\bibitem{Gupta:2021ahb}
R.~Gupta {\em et~al.},
\newblock Phys. Rev. Lett. {\bf 127}, 242002 (2021), 2105.12095.

\bibitem{Hoferichter:2023ptl}
M.~Hoferichter, J.~R. de~Elvira, B.~Kubis, and U.-G. Mei\ss{}ner,
\newblock Phys. Lett. B {\bf 843}, 138001 (2023), 2305.07045.

\bibitem{Borsanyi:2020bpd}
S.~Borsanyi {\em et~al.},
\newblock (2020), 2007.03319.

\bibitem{Krebs:2020plh}
H.~Krebs, E.~Epelbaum, and U.-G. Mei\ss{}ner,
\newblock Eur. Phys. J. A {\bf 56}, 240 (2020), 2005.07433.

\bibitem{Krebs:2019aka}
H.~Krebs, E.~Epelbaum, and U.-G. Mei\ss{}ner,
\newblock Few Body Syst. {\bf 60}, 31 (2019), 1902.06839.

\bibitem{Nogga:2001cz}
A.~Nogga, H.~Kamada, W.~Gl{\"o}ckle, and B.~R. Barrett,
\newblock Phys. Rev. C {\bf 65}, 054003 (2002), nucl-th/0112026.

\bibitem{Wiringa:1994wb}
R.~B. Wiringa, V.~G.~J. Stoks, and R.~Schiavilla,
\newblock Phys. Rev. C {\bf 51}, 38 (1995), nucl-th/9408016.

\bibitem{Machleidt:2000ge}
R.~Machleidt,
\newblock Phys. Rev. C {\bf 63}, 024001 (2001), nucl-th/0006014.

\bibitem{Epelbaum:2014sza}
E.~Epelbaum, H.~Krebs, and U.-G. Mei\ss{}ner,
\newblock Phys. Rev. Lett. {\bf 115}, 122301 (2015), 1412.4623.

\bibitem{Reinert:2017usi}
P.~Reinert, H.~Krebs, and E.~Epelbaum,
\newblock Eur. Phys. J. A {\bf 54}, 86 (2018), 1711.08821.

\bibitem{Epelbaum:2014efa}
E.~Epelbaum, H.~Krebs, and U.-G. Mei\ss{}ner,
\newblock Eur. Phys. J. A {\bf 51}, 53 (2015), 1412.0142.

\bibitem{Krebs:2020pii}
H.~Krebs,
\newblock Eur. Phys. J. A {\bf 56}, 234 (2020), 2008.00974.

\bibitem{Filin:2020tcs}
A.~A. Filin {\em et~al.},
\newblock Phys. Rev. C {\bf 103}, 024313 (2021), 2009.08911.

\bibitem{Furnstahl:2015rha}
R.~J. Furnstahl, N.~Klco, D.~R. Phillips, and S.~Wesolowski,
\newblock Phys. Rev. C {\bf 92}, 024005 (2015), 1506.01343.

\bibitem{Melendez:2019izc}
J.~A. Melendez, R.~J. Furnstahl, D.~R. Phillips, M.~T. Pratola, and
  S.~Wesolowski,
\newblock Phys. Rev. C {\bf 100}, 044001 (2019), 1904.10581.

\bibitem{Griesshammer:2020ufp}
H.~W. Grie\ss{}hammer, J.~A. McGovern, A.~Nogga, and D.~R. Phillips,
\newblock Few Body Syst. {\bf 61}, 48 (2020), 2005.12207.

\bibitem{Nogga:2023nucdens}
A.~Nogga,
\newblock {NUCDENS}: {A} package to access momentum space densities for light
  nuclei.; v1.1.0, 2023.

\bibitem{Smith:2006ym}
M.~C. Smith {\em et~al.},
\newblock Mon. Not. Roy. Astron. Soc. {\bf 379}, 755 (2007), astro-ph/0611671.

\bibitem{Cirigliano:2020dmx}
V.~Cirigliano, W.~Dekens, J.~de~Vries, M.~Hoferichter, and E.~Mereghetti,
\newblock Phys. Rev. Lett. {\bf 126}, 172002 (2021), 2012.11602.

\bibitem{Wirth:2021pij}
R.~Wirth, J.~M. Yao, and H.~Hergert,
\newblock Phys. Rev. Lett. {\bf 127}, 242502 (2021), 2105.05415.

\bibitem{Jokiniemi:2021qqv}
L.~Jokiniemi, P.~Soriano, and J.~Men\'endez,
\newblock Phys. Lett. B {\bf 823}, 136720 (2021), 2107.13354.

\bibitem{Hoferichter:2018acd}
M.~Hoferichter, P.~Klos, J.~Men\'endez, and A.~Schwenk,
\newblock Phys. Rev. D {\bf 99}, 055031 (2019), 1812.05617.

\bibitem{Chang:2017eiq}
NPLQCD, E.~Chang {\em et~al.},
\newblock Phys. Rev. Lett. {\bf 120}, 152002 (2018), 1712.03221.

\bibitem{Friar:1979zz}
J.~L. Friar,
\newblock Phys. Rev. C {\bf 20}, 325 (1979).

\bibitem{Liebig:2015kwa}
S.~Liebig, U.-G. Mei\ss{}ner, and A.~Nogga,
\newblock Eur. Phys. J. A {\bf 52}, 103 (2016), 1510.06070.

\bibitem{Le:2020zdu}
H.~Le, J.~Haidenbauer, U.-G. Mei\ss{}ner, and A.~Nogga,
\newblock Eur. Phys. J. A {\bf 56}, 301 (2020), 2008.11565.

\end{thebibliography}

\end{document}